\documentclass[%
 reprint,
 superscriptaddress, 
 amsmath,amssymb,
 aps,
 pra,
 floatfix,
]{revtex4-2}

\usepackage{graphicx}
\usepackage{dcolumn}
\usepackage{bm}
\usepackage{dsfont}
\usepackage{xcolor}
\usepackage{enumitem}
\usepackage{multirow}
\usepackage{hyperref}
\usepackage{array}
\usepackage{float}
\usepackage{csquotes}
\MakeOuterQuote{"}

\usepackage{tikz}
\usepackage{amsthm}
\usepackage{amssymb}

\theoremstyle{definition}

\newcommand{\B}{\mathcal{B}}

\begin{document}

\title{Faster-than-Clifford Simulations of Entanglement Purification Circuits and Their Full-stack Optimization}

\author{Vaishnavi L. Addala}
\thanks{These two authors contributed equally}
\affiliation{Department of Electrical Engineering and Computer Science, Massachusetts Institute of Technology,
77 Massachusetts Avenue, Cambridge, Massachusetts 02139, USA}

\author{Shu Ge}
\thanks{These two authors contributed equally}
\affiliation{Department of Electrical Engineering and Computer Science, Massachusetts Institute of Technology,
77 Massachusetts Avenue, Cambridge, Massachusetts 02139, USA}

\author{Stefan Krastanov}
\email{skrastanov@umass.edu}
\affiliation{Department of Electrical Engineering and Computer Science, Massachusetts Institute of Technology,
77 Massachusetts Avenue, Cambridge, Massachusetts 02139, USA}
\affiliation{ College of Information and Computer Sciences, University of Massachusetts Amherst, 140 Governors Drive
Amherst, Massachusetts 01003, USA}

\date{\today}

\begin{abstract}
Quantum Entanglement is a fundamentally important resource in Quantum Information Science; however, generating it in practice is plagued by noise and decoherence, limiting its utility. Entanglement distillation and forward error correction are the tools we employ to combat this noise, but designing the best distillation and error correction circuits that function well, especially on today's imperfect hardware, is still challenging. Here, we develop a simulation algorithm for distillation circuits with gate-simulation complexity of $\mathcal{O}(1)$ steps, providing for drastically faster modeling compared to $\mathcal{O}(N)$ Clifford simulators or $\mathcal{O}(2^N)$ wavefunction simulators over $N$ qubits.

This new simulator made it possible to not only model but also optimize practically interesting purification circuits. It enabled us to use a simple discrete optimization algorithm to design purification circuits from $n$ raw Bell pairs to $k$ purified pairs and study the use of these circuits in the teleportation of logical qubits in second-generation quantum repeaters. The resulting purification circuits are the best-known purification circuits for finite-size noisy hardware and can be fine-tuned for specific hardware error models. Furthermore, we design purification circuits that shape the correlations of errors in the purified pairs such that the performance of the error-correcting code used in teleportation or other higher-level protocols is greatly improved. Our approach of optimizing multiple layers of the networking stack, both the low-level entanglement purification, and the forward error correction on top of it, are shown to be indispensable for the design of high-performance second-generation quantum repeaters.
\end{abstract}

\maketitle


Entanglement, a fundamental resource in quantum information theory, allows for "stronger" correlations than what is possible classically. This resource is pivotal for numerous foundational quantum technologies including quantum key distribution, superdense coding, state and gate teleportation, and error correction, among others. However, the practical generation of entangled quantum resources, specifically Bell pairs, on current hardware suffers from significant noise, with error rates on the order of 10\% in platforms as diverse as trapped ions \cite{hucul2015modular, moehring2007entanglement, krutyanskiy2023entanglement}, color centers \cite{pfaff2013demonstration, hensen2015loophole,hermans2023entangling}, neutral atoms or ensembles \cite{ritter2012elementary,yang2022sequential,van2022entangling, yu2020entanglement,luo2022postselected}, microwave qubits \cite{narla2016robust} and others \cite{rakonjac2023transmission,jiang2022quantum}. This infidelity poses challenges for the deployment of quantum resources in a network of any scale, whether metropolitan, local, or "on chip", necessitating the development of circuits for entanglement "purification".

Entanglement purification protocols aim to improve the fidelity of entangled resources through local operations and classical communications (LOCC). Many such protocols have been developed \cite{deutsch1996quantum,bennett1996purification,dur1999quantum,dur2007entanglement,fujii2009entanglement, nickerson2013topological,nickerson2014freely,nigmatullin2016minimally,krastanov2019optimized}, with well established optimal results for the "asymptotic" regime of perfect gates and large memories. However, many of these results do not retain their optimality on imperfect hardware and recent works have started focusing on optimizing these protocols for finite-size noisy hardware either by using bespoke hand-crafted circuits \cite{nickerson2013topological,nickerson2014freely,nigmatullin2016minimally}, or by employing optimization and machine learning techniques \cite{krastanov2019optimized}, or through exhaustive enumeration \cite{jansen2022enumerating,goodenough2023near}.

Challenges persist in the design and optimization of these protocols, owing to the need to balance high output fidelity with experimental feasibility. We still do not know what "the best" purification circuit is when the size of the hardware register is constrained and when the local operations are imperfect. Moreover, the machine optimization of purification circuits becomes limited by the need for sufficiently fast simulation.

We design a new simulation method, with much lower computational complexity than alternatives, that permits us to run optimization heuristics on much larger circuits. This approach provides the first conclusive answer to how to perform multi-pair purification optimized for the particularities of real hardware.

Moreover, we go beyond considering purification in isolation and study its use in full-stack protocols. We show how important it is to be careful with the choice of figure of merit when designing entanglement purification. For instance, the definition of fidelity employed in the vast majority of entanglement purification literature is actively detrimental if the purified pairs are to be used in state teleportation protocols (one of the most common applications), leading to the worst performance of all figures of merit considered in this work. As discussed in the main text, this is due to the multi-qubit correlated errors that are not constrained by the usual definition of fidelity. 

This type of co-design of purification informed by the lower levels of the stack (optimizing for the specific hardware error model) and higher-level applications (e.g., error-corrected teleportation) is crucial for the development of quantum informational technologies.

This paper is structured as follows: In section~\ref{sec:purification_basics}, we give an overview of entanglement purification techniques, how noise can be modeled, and our approach for circuit optimization. In section~\ref{sec:entanglement_optimization}, we discuss the performance of the generated protocols, compare them to known protocols, and show they significantly outperform alternative methods, especially when co-designed with the rest of the technology stack. The heavy optimizations necessary for this achievement are only possible thanks to our new faster-than-Clifford purification simulation technique, which is discussed at length in section~\ref{sec:fast_simulation}. We end this paper with concluding remarks on applicability to more general quantum networking problems and a number of problems that now become much easier to tackle.

\section{Basics of Entanglement Purification \label{sec:purification_basics}}

Typically, in an entanglement purification protocol, a number of Bell pairs are sacrificed in order to perform a nonlocal measurement on another set of Bell pairs. 
In an $n$ to $k$ entanglement purification procedure, two parties start out by sharing $n$ raw Bell pairs and sacrifice $n-k$ of them by applying local operations and measurements and communicating classically in order to obtain $k$ pairs of higher fidelity. In coming up with an optimized protocol, we must take into consideration the errors in the raw Bell pairs and the errors from the procedure itself. As we increase $n$, there are exponentially many possible circuits to evaluate if we are to do a brute force search. We employ a genetic algorithm~\cite{krastanov2019optimized} for the search and restrict the permitted gates to the "good subset" of Clifford gates which only perform permutations in the Bell basis (which also enables the use of a much more efficient simulation algorithm we detail in a later section).

The search algorithm starts with a population of circuits randomly created with the allowed operations. At each step, mutations and offspring are added to the population and only the best-performing circuits will be kept based on the cost function. The search ends when the performance converges amongst the population. For the majority of our text, we use the cost function $F_\textrm{out}$ described at the end of the section. Most searches take less than a minute thanks to the new simulation algorithm we have developed (presented in the second half of this text).

The allowed operations within the circuits are mirrored CNOTs, coincidence/anti-coincidence measurements, and local 1-qubit Clifford gates that perform a permutation in the Bell basis of a single pair. In a coincidence measurement, the two parties involved (Alice and Bob) throw away their Bell pairs and restart the procedure if their measurement results differ. They restart if their results are the same in the case of an anti-coincidence. Note that Alice and Bob communicate classically so they can compare the results of their measurement. CNOTs propagate some errors between the qubits involved. Based on this, when a CNOT is followed by a coincidence measurement, we can detect some errors on the pair that we have not measured. With this information, we can choose to keep that pair or discard it and restart the procedure.
Another way of thinking about these circuits is that they perform permutations in the Bell basis. Here, we can interpret finding the best circuit as finding the best set of permutations followed by the coincidence measurements which cause the probability of the pairs being pure to be maximized.

\begin{figure}
    \centering
    \includegraphics[width=\linewidth]{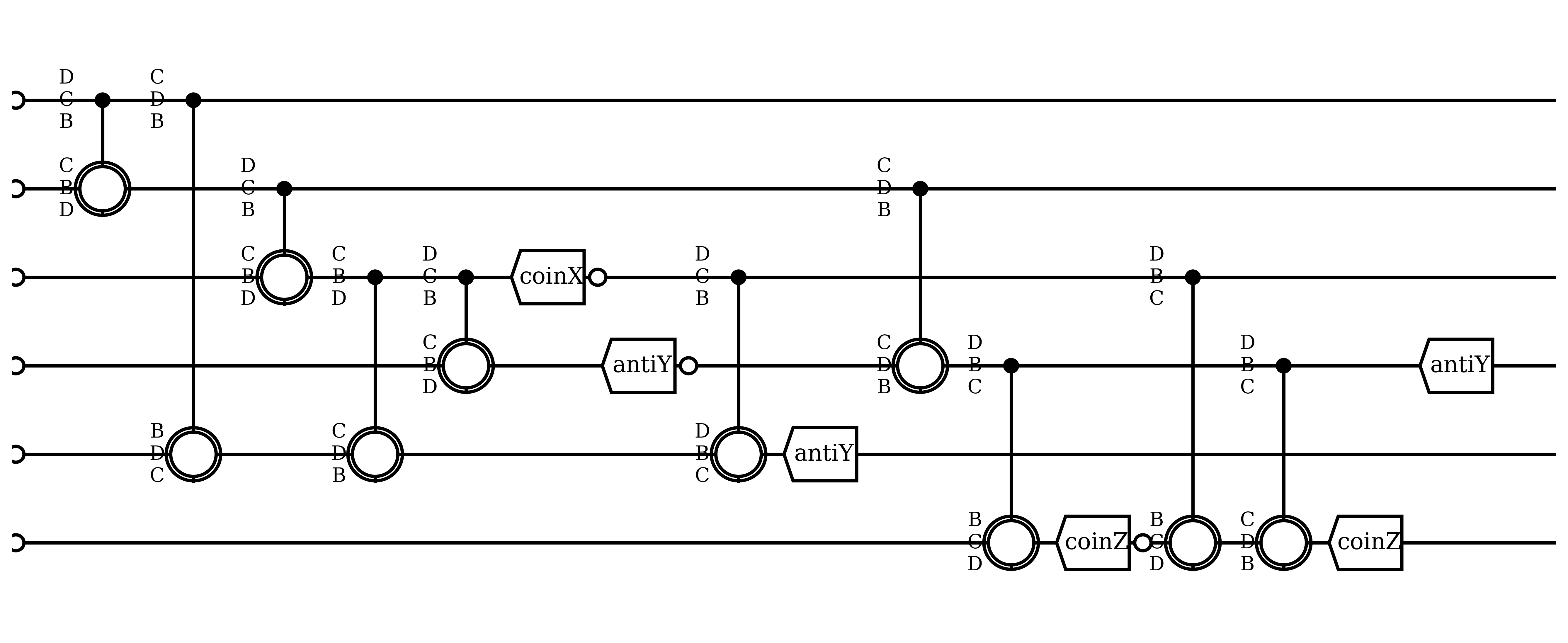}
    \caption{\textbf{Example optimized purification circuit}. This circuit produces $k=3$ purified Bell pairs using registers of finite number $r=6$ and $n=9$ raw Bell pairs. The 9 small open circles seen at the beginning of the circuit and after intermediary measurements signify the generation of a raw Bell pair in that register. Note that we are only showing Alice's circuit and that Bob performs the same operations on his circuit. CNOTs are preceded by a permutation in the Bell basis and these are together shown as open circles with the permutation written out to the left. The measurements are labeled with whether they are coincidence or anti-coincidence and in which Pauli eigenbasis they are performed.}
    \label{fig:9to3}
\end{figure}

There are two steps where we model error: in the generation of a raw Bell pair and in 2-qubit gates and measurements. We can represent any set of Bell pairs with a density matrix diagonal in the Bell basis, however for initial raw pairs we typically pick the Werner state $\rho_0 = F_\textrm{in}|A\rangle \langle A|+\frac{1-F_\textrm{in}}{3}(|B\rangle \langle B|+|C\rangle \langle C|+|D\rangle \langle D|)$, where the different Bell states are: 
\begin{center} $|A\rangle = \frac{|00\rangle + |11\rangle}{\sqrt{2}}\hspace{2em} 
|B\rangle = \frac{|01\rangle - |10\rangle}{\sqrt{2}}$\end{center}
\begin{center}
$|C\rangle = \frac{|01\rangle + |10\rangle}{\sqrt{2}}\hspace{2em} 
|D\rangle = \frac{|00\rangle - |11\rangle}{\sqrt{2}} $\end{center}
$F_\textrm{in}$ is the fidelity of a raw Bell pair while $1-F_\textrm{in}$ can be thought of as the overall error or infidelity.

For 2-qubit gates, we assume the gate acts correctly with probability $p_2$ and completely depolarizes both pairs involved with probability $1-p_2$. However, if more information about the types of biased errors on any particular hardware system is known, our framework can be used to model those errors instead (as long as they can be represented by a Pauli channel). For most of the numerical experiments in this text, following typical current hardware parameters, the error rate of raw Bell pairs is on the order of 10\%, the error rate of 2-qubit gates is on the order of 1\%, and we do not consider 1-qubit gates since their error rates are much smaller.

For measurement errors, we assume that with probability $\eta$ the correct results of the projection are returned, with probability $1-\eta$ the incorrect result is returned. When not specified, we assume $\eta=p_2$. The allowed measurements are coincidences in the X and Z Pauli eigenbases and anti-coincidence in the Y eigenbasis, which each select for the state $|A\rangle$ and for one other Bell state. In other words, when these measurements fail (differing results for coincidence and same results for anti-coincidence), Alice and Bob have detected an error in one of our Bell pairs so they restart the procedure.

Lastly, before exploring the zoo of highly optimized purification circuits we have discovered, we need to discuss a variety of optimization targets and cost functions available to us. As discussed later in this text, this choice can be crucial for the overall performance of a quantum network. For the majority of the presented results we consider a quantity denoted $F_\textrm{out}$ which we call "average marginal fidelity". The states we consider have diagonal density matrix in the multi-pair Bell basis, and as such can be interpreted as a discrete probability distribution over multiple variables -- each variable corresponds to one of the Bell pairs and each of the discrete possibilities is whether we have an A, B, C, or D state. The marginal probability to have the $i^\textrm{th}$ pair in state A, averaged over the indices $i$ corresponding to purified pairs is $F_\textrm{out}$. In a typical "physical" notation that is $$F_\textrm{out}=\frac{1}{k}\sum_{i\in1..k}\langle A|\operatorname{ptrace}_{\{i\}}(\rho)|A\rangle,$$
where $k$ is the number of purified pairs, $\rho$ is the final multi-pair density matrix, and $\operatorname{ptrace}_S$ is partial trace over all subsystems not in the set $S$. We stress that directly computing $F_\textrm{out}$ in our simulations is more straightforward than the expression given above implies, thanks to the data structures we use, as discussed later in this text.

In a few specific contexts, we also consider the more typical figure of merit $F_A$ corresponding to all purified Bell pairs being in state A: $$F_A = \langle A^{\otimes k}|\operatorname{ptrace}_{\{1..k\}}(\rho)|A^{\otimes k}\rangle.$$ We show that this very common choice of figure of merit can be detrimental, and provide better alternatives to it.

Another metric we consider is success probability. This is the probability of the circuit having a successful run, which corresponds to Alice and Bob never restarting their procedure based on their measurement results. This is a worst-case bound, as in many cases it is feasible for Alice and Bob to restart only a small part of their circuit without starting from scratch.

\section{Optimized Purification Circuits\label{sec:entanglement_optimization}}

\begin{figure}
    \centering
    \includegraphics[width=\linewidth]{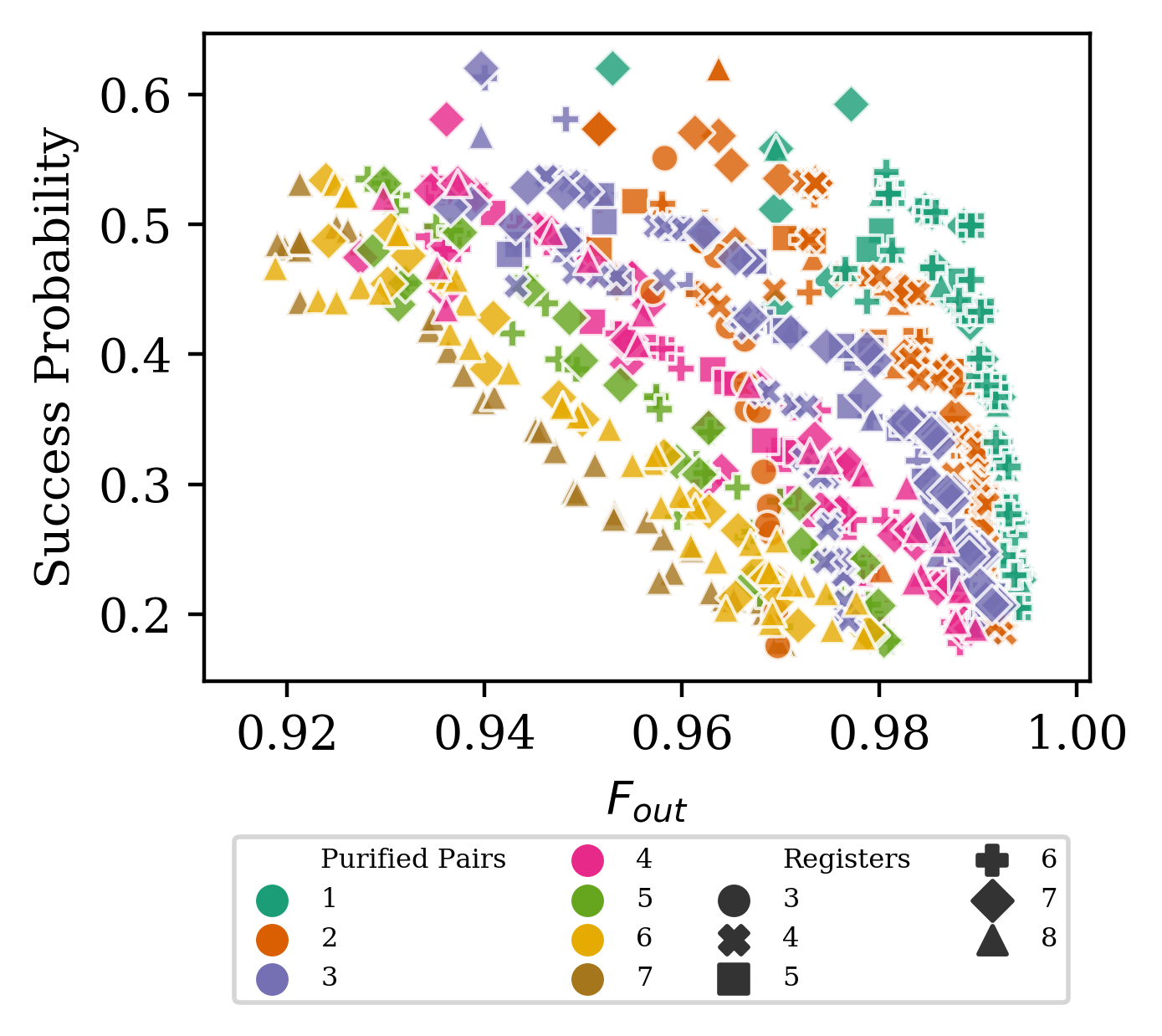}
    \caption{\textbf{A population of the highest performing generated circuits}. The horizontal axis is $F_\textrm{out}$ - the fidelity of the purified pairs, the vertical axis is how likely the procedure is to succeed, the colors show $k$ - how many pairs the circuits purify to, and the shapes show $r$ - the register width of the circuit. The optimizations were done at $F_\textrm{in}=0.9,p_2=0.99$ without restricting the number of raw Bell pairs ($n$) or circuit length (visualized in the following figures). The most appropriate circuit to choose from this population would depend on hardware constraints and application-level goals as described in the main text. Importantly, the optimization can easily be redone to fine-tune the circuits for particular hardware. Some general trends we see are: increasing the number of preserved purified pairs leads to lower fidelity, while more sacrificed pairs lead to higher fidelity; increased fidelity comes with decreased success probability; increasing register width tends to increase fidelity but the increase gets saturated quickly at $r=n+2$~\cite{krastanov2019optimized,fujii2009entanglement} -- easily seen in the $k=2$ and $k=3$ cases at various $r$ and in the $r=7$ and $r=8$ cases at various $k$. These trends are explored in more detail in Fig.~\ref{fig:Nto2} where the dependence on $n$ is elucidated as well.}
    \label{fig:CircuitsPopulation}
\end{figure}

\begin{figure}
    \centering
    \includegraphics[width=\linewidth]{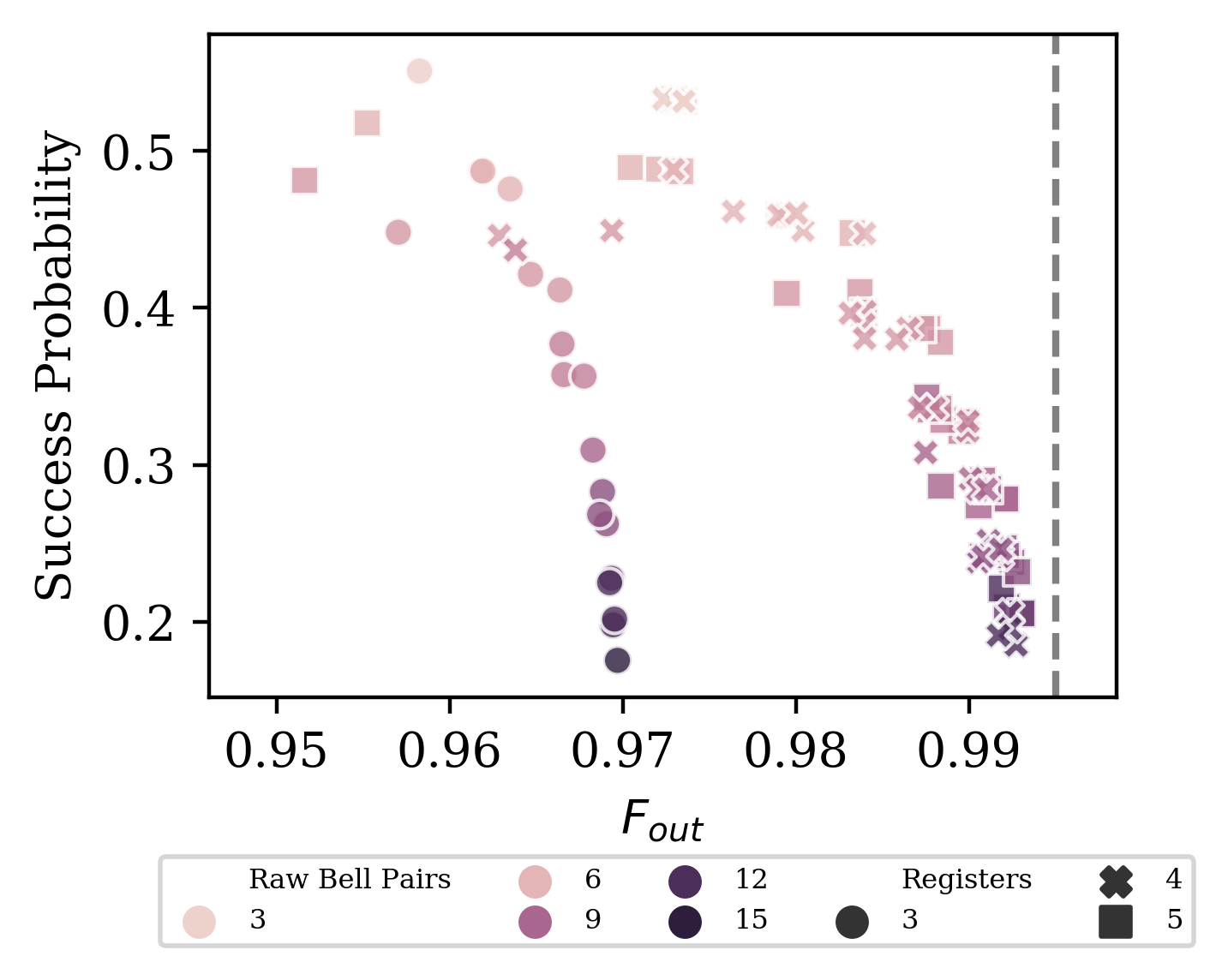}
    \caption{\textbf{Operational error bound and minimum register width to approach it}. Each data point is an $n$-to-2 circuit evaluated at $F_\textrm{in}=0.9$, $p_2=0.99$. Axes as in Fig.~\ref{fig:CircuitsPopulation} and the color indicates the number of raw Bell pairs used. We see that increasing the number of raw pairs sacrificed corresponds to higher $F_\textrm{out}$ since we are able to perform more rounds of error detection, which however leads to decreased single-shot success probability. There is a jump in $F_\textrm{out}$ as the register width grows from three to four due to the fact that four registers are the minimum needed to perform "double selection"~\cite{fujii2009entanglement}. Once this minimum is met, the circuits approach the upper bound on fidelity, shown as a vertical dashed line which is set by the error of the last operations performed~\cite{krastanov2019optimized}. }
    \label{fig:Nto2}
\end{figure}

We explored a number of situations where our optimization approach, enabled by the new method for fast simulations, provides state-of-the-art $n$-to-$k$ purification circuits. First, before studying specific important cases, we will present a survey over a wide range of circuit parameters, showcasing both the ease with which our technique can be employed and a number of important trade-offs in the design of purification circuits. For illustration purposes, Fig.~\ref{fig:9to3} presents a typical discovered (optimized) purification circuit.

For an initial parameter sweep population study, we consider $n$-to-$k$ purification circuits, on registers of varying size. The consideration of register size limitations is quite crucial, as many of the "good" well-known recursive purification circuits from the literature are impossible to implement on small-size registers. We denote $n$ the number of initial raw pairs, $k$ the number of preserved pairs, and $r$ the size of the register. With register reuse $r<n$ is permitted, but naturally we need $k<r$ for any purification to be possible. In this parameter sweep we consider $r$ up to 8, $k$ up to 7, and we do not bound $n$. The noise of the two-qubit gates is also taken into account: a crucial limiting factor very rarely considered in most literature on the topic. The resulting circuits are summarized in Fig.~\ref{fig:CircuitsPopulation}. One of the more important features that can be observed is the need for a register wide enough to contain more than two sacrificial pairs -- otherwise the error detection capabilities of the circuit are restricted as previously observed in~\cite{krastanov2019optimized,fujii2009entanglement}. There is a trade-off between the probability of success and fidelity, which is also dependent on the number of sacrificial pairs (more thoroughly investigated in Fig.~\ref{fig:Nto2}). That figure also shows an asymptote for the achievable fidelity, present due to gate noise (in the absence of scalable large registers permitting error-correcting gates).

With the faster simulation methods we have developed, such a population study can be performed in minutes on commodity classical hardware. For pedagogical purposes, we picked fairly generic parameters, but any hardware developer can easily redo the simulations, taking into account biased noise typical for their platform, memory errors, or finite wait times between raw pair generation. This co-design process provides significant gains in performance, compared to applying simple generic purification circuits as seen from comparisons in the following paragraphs.

Moreover, colleagues studying methods for an exhaustive enumeration of possible purification circuits have graciously provided comparisons between our optimization techniques and a few "known best circuits in the absence of gate noise"~\cite{goodenough2023near}, giving additional proof of the high performance of our circuits. Below we continue with comparisons to typical small purification circuits.

\subsection{Comparison to Other Methods}

\begin{figure}
    \centering
    \includegraphics[width=\linewidth]{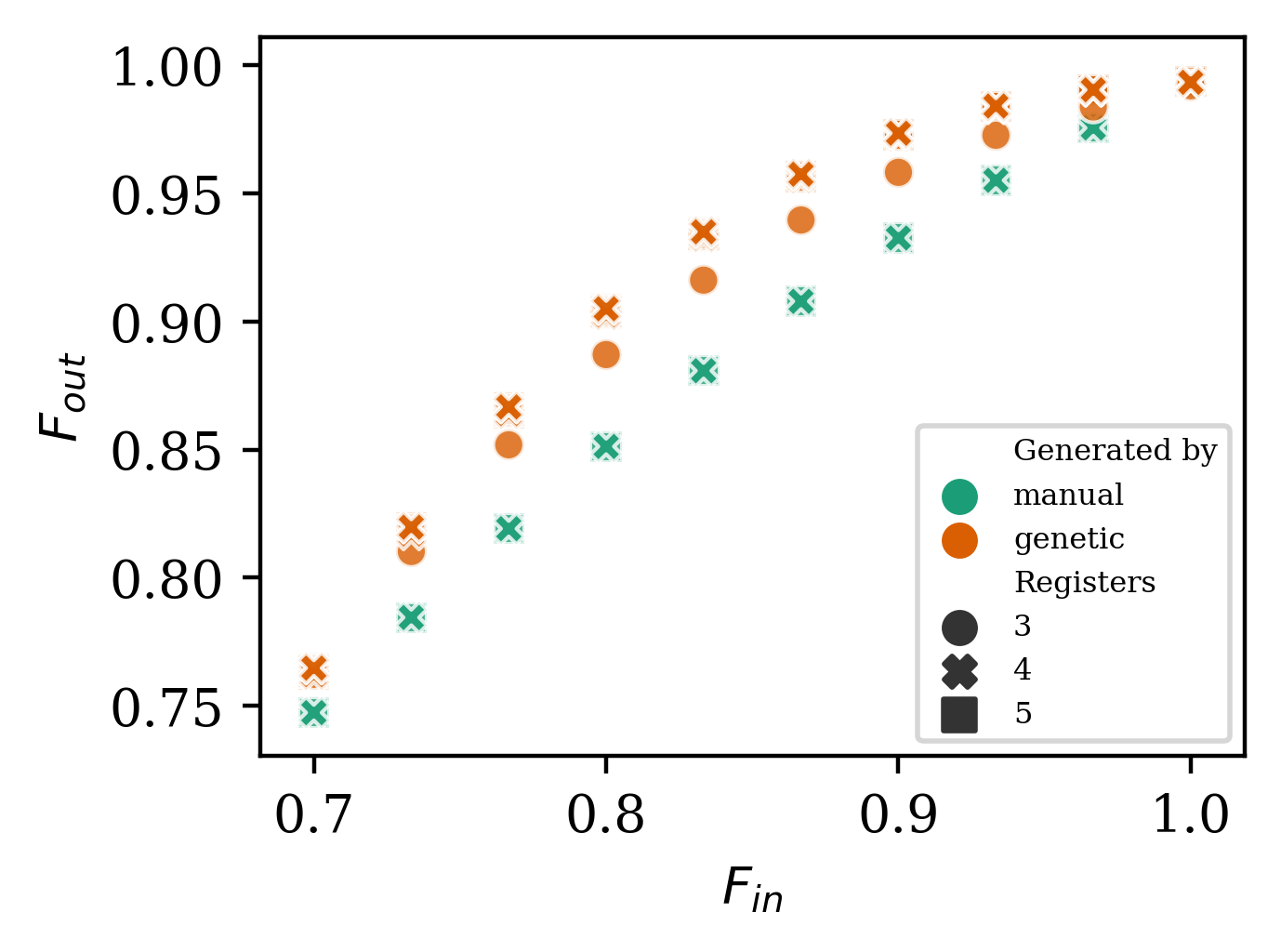}
    \caption{\textbf{Comparison to using standard building block circuits}. The plot is $F_\textrm{in}$ versus $F_\textrm{out}$ and the colors represent whether the circuit is manually created from standard sub-circuits found in the literature or generated by our optimizer. Examples of these circuits can be found in Fig.~\ref{fig:5to2Circuits}. Our optimized circuits outperform circuits built from single and double selections sub-circuits in a wide range of raw Bell pair fidelities and can have smaller register widths. The circuits are evaluated at $p_2=0.99$.}
\label{fig:5to2}
\end{figure}

\begin{figure}
    \centering
    \includegraphics[width=\linewidth]{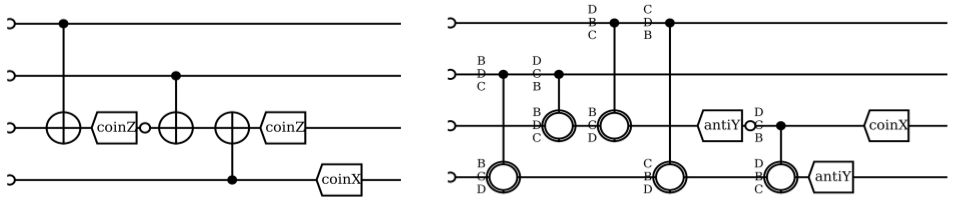}
    \caption{\textbf{5-to-2 "manually assembled" circuit and optimized circuit}. The circuit on the left is manually created, with a single selection circuit followed by the most appropriate double selection. On the right is an example of a circuit found by the optimizer.}
    \label{fig:5to2Circuits}
\end{figure}

\begin{figure}
    \centering
    \includegraphics[width=\linewidth]{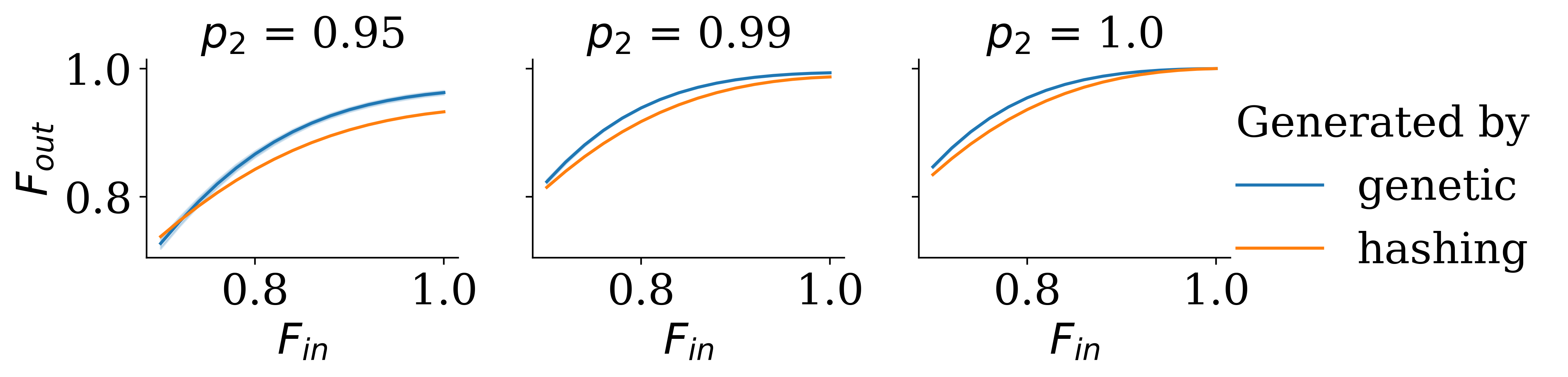}
    \caption{\textbf{Comparison to truncated hashing protocol}. The plot is $F_\textrm{in}$ versus $F_\textrm{out}$, the blue is our circuits while the orange is the truncated hashing protocol circuit, and we have three sub-graphs at different gate fidelities. The optimized circuits are able to outperform the truncated hashing protocol at various hardware parameters after being optimized for $F_\textrm{in}=0.9,p_2=0.99$.}
    \label{fig:Hashing}
\end{figure}

There are very few non-asymptotic designs for $n$-to-$k$ circuits in the literature, virtually none that are designed with gate (a.k.a. "operational") noise constraints in mind. Even "information-theoretically perfect" methods like the hashing protocol~\cite{bennett1996mixed} fall short when applied to finite circuits or when gate noise is present. Below we compare our optimized circuits to straightforward concatenations of known protocols and truncated versions of the otherwise asymptotic "perfect" hashing protocol, showing we significantly outperform both approaches.

Consider the well-known single- and double-selection~\cite{nickerson2013topological,fujii2009entanglement} circuits. We compare against manually generated circuits built from these sub-circuits in Fig.~\ref{fig:5to2}. The increase in performance of our optimized circuits comes from being able to entangle more Bell pairs before measuring and detecting errors, which we are confined to in a modular setting. For the same reason, our technique can provide the same final fidelity at much lower hardware requirements.

On the other end of the spectrum, consider the hashing protocol which early on was designed to saturate information-theory bounds about purification performance (assuming asymptotically many pairs in unbounded register with perfect gates). We next evaluate our methods against the hashing protocol when it is truncated to fit on finite hardware. The truncated hashing protocol works best when purifying to and from powers of two Bell pairs, so for that reason, we show 8 to 2 circuits in Fig.~\ref{fig:Hashing}. Similarly to the previous comparison, we significantly outperform hashing, especially with noisy gates.

Both of these alternative methods we compare against are in settings they are not optimized for, but nonetheless, they are the only techniques otherwise available. It is worth noting that there is a correspondence between $n$-to-$k$ purification protocols and $[[n,k]]$ error correcting codes, but similarly to the troubles with the hashing protocol, studying that correspondence does not take into account the finite size of available registers, the gate noise in those registers, and the exact layout of the gates in such a circuit.

More generally, optimization methods like ours permit co-design of both the abstract purification protocol and the circuit compilation itself, which is extremely valuable if the hardware has some limited qubit-connectivity topology. E.g. if our qubits permit only nearest-neighbor gates, it is not enough to know the best purification protocol, but rather the best protocol given the limited qubit-connectivity. Instead of solving the purification protocol design and the circuit compilation problem separately, optimizers like ours solve both problems together, providing for much higher performance compared to concatenating two independent solutions.

\subsection{Application to Error Correction and Quantum Repeaters}

Here we pursue a more sophisticated example of such co-design, where we do not simply maximize the performance of a single purification circuit, but rather the performance of the entire application stack build on top of the purification circuit. Consider the use of Bell pairs in quantum repeaters for instance. In 2nd and 3rd generation quantum repeaters~\cite{muralidharan2016optimal}, teleportation is followed by error correction, and this is a natural application for our purified circuits. The process would look as follows:
\begin{enumerate}
    \item Purification of $n$ raw Bell pairs into $k$ higher quality pairs
    \item Teleportation of logical qubit(s) encoded in $k$ physical qubits
    \item Error correction on the logical qubit(s)
\end{enumerate}

We can then determine the quality of the procedure by how likely it is that the logical qubit is in its original state after teleportation and error correction. This overall probability we will call the logical fidelity $F_L$. Note that in calculating $F_L$, we assume the local gates for the error correction procedures are noiseless, in order to focus on the purification step (but including noisy gates for the error correction would be straightforward). We show results when using the 5-qubit~\cite{bennett1996mixed, laflamme1996perfect} and a [[11, 1, 5]] error correction codes (the smallest to correct for two-qubit errors~\cite{gottesman_thesis}), but the technique is generally applicable to any stabilizer code.

\begin{figure}
    \centering
    \includegraphics[width=\linewidth]{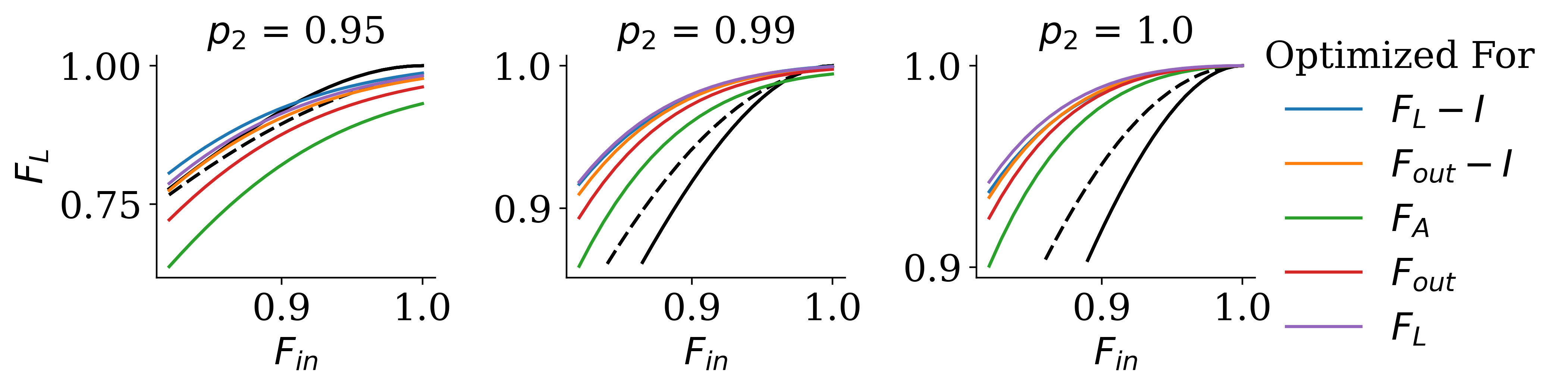}
    \caption{\textbf{Fidelity of logical teleportation using 10-to-5 purification and the 5 qubit code}. The plot is $F_\textrm{in}$ versus the fidelity of the logical teleported qubit after error correction $F_L$, while the colors signify what the circuits were optimized for. The solid black line shows the performance of using unpurified pairs for teleportation, and the dashed black line shows the performance when using the typical 2-to-1 single selection purification protocol (equivalent to truncating the hashing protocol). Surprisingly, we see that optimizing for purification fidelity ($F_A$ or $F_\textrm{out}$) leads to significantly lower overall performance. As discussed in the main text, this is due to not constraining the long tail of correlated multi-qubit errors. All optimizations were done at raw Bell fidelity $F_\textrm{in}=0.9$ and local gate fidelity $p_2=0.99$, but we plot the performance at various $p_2$. If the $p_2$ value is known for a given hardware, the circuit should be optimized for that value. Further exploration of various circuit parameters, like circuit length and width is presented in the appendix.}
\label{fig:Teleportation}
\end{figure}

\begin{figure}
    \centering
    \includegraphics[width=\linewidth]{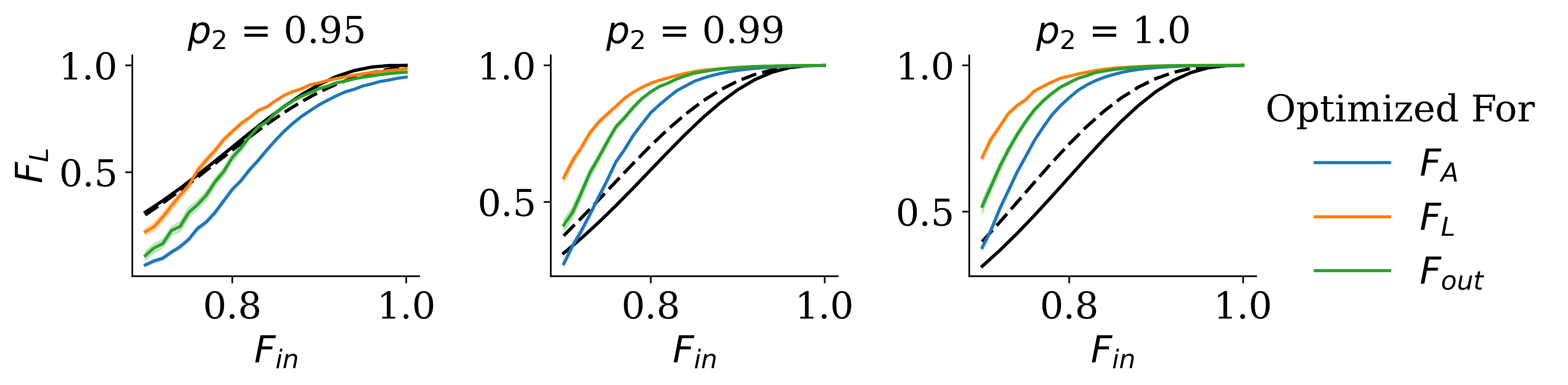}
    \caption{\textbf{Fidelity of logical teleportation using the [[11, 1, 5]] code}. The plot is $F_\textrm{in}$ versus $F_L$, the colors signify what the circuits were optimized for, the solid black line shows the performance of using unpurified pairs for teleportation, and the dashed black line shows the performance when using the 2 to 1 single selection protocol to purify pairs. We see similar trends as we saw with the 5-qubit code in Fig.~\ref{fig:Teleportation}, which shows how our method can scale for larger codes and code distances. These circuits use 14 registers and are evaluated with 100,000 Monte Carlo runs.}
\label{fig:TeleportationLargerCode}
\end{figure}

The overall theme behind the results is that by doing full-stack optimization instead of optimizing only for entanglement purification fidelity, we consistently obtain circuits with significantly better performance for the two codes we have considered as seen in Fig.~\ref{fig:Teleportation} and Fig.~\ref{fig:TeleportationLargerCode}. The exact circuits are specific for the network error model, gate error rate, error correction code, and other constraints, thus we do not plot them here, however, they can be seen in the archived software repository or regenerated within minutes thanks to the provided software (while also being optimized for the particular hardware parameters that matter to the downstream user).

We explored three main cost functions in our optimizer. The first is $F_\textrm{out}$, which is the average marginal fidelity of each of the purified pairs (the metric we use throughout most of the paper as a proxy for quality of purification). The second is $F_A$, which is the probability that all the purified pairs are to be Bell pair A, i.e., the overlap of the final purified state with the state $|A\rangle^{\otimes k}$ -- this is another very common measure of quality of purification. The third is $F_L$, which is the probability of no logical error after teleportation (i.e., the probability for fewer than $d/2$ errors in the purification protocol, where $d$ is the code distance). This third cost function is equivalent to the fidelity of the teleported logical qubit.

The surprising result we observe is that optimizing for high-quality purification does not result in high-quality overall protocol performance, underscoring the importance of co-design. The intuitive reason behind this effect is that a cost function that focuses on high-quality purification does not constrain the long tail of the probability distribution over possible errors, which is crucial to the performance of an error-correcting code. I.e., $F_A$ and to a lesser extent $F_\textrm{out}$ penalize low-weight and high-weight errors equally. Conversely, one can interpret the $F_L$ cost function as forbidding multi-qubit errors while permitting correctable errors, i.e., errors with a weight smaller than $d/2$ where $d$ is the distance of the code layered on top of the purification. Other measures of error correlation were also considered as cost functions and discussed in the appendix.

\section{Faster Simulation of Entanglement Circuits \label{sec:fast_simulation}}

We now introduce the faster simulation algorithm that made much of the optimization work possible. Typical entanglement purification circuits are Clifford circuits used on stabilizer states, for which there are known polynomial algorithms \cite{gottesman1998heisenberg, aaronson2004improved}. The well-known Stabilizer tableaux representation typically takes $\mathcal{O}(N^2)$ space complexity, where $N$ is the number of qubits. Applying a single gate takes $\mathcal{O}(N)$ time, and a measurement takes $\mathcal{O}(N^2)$. The graph state formalism is even faster (for sparse tableaux), with $\mathcal{O}(N\log N)$ space complexity~\cite{anders06}.

In order to design a more effective algorithm to simulate purification circuits, we propose a more efficient representation of Bell states and purification circuits. We limit ourselves only to multi-pair mixed state entanglement purification. We consider states whose density matrix is diagonal in the Bell basis $\{A,B,C,D\}^{\otimes n}$ ($n$ is the number of Bell pairs). 

\begin{align}
    A&\propto|00\rangle + |11\rangle \sim 
    \begin{bmatrix}
        + &XX\\
        + &ZZ
    \end{bmatrix}\\
    B&\propto|01\rangle - |10\rangle \sim 
    \begin{bmatrix}
        - &XX\\
        - &ZZ
    \end{bmatrix}\\
    C&\propto|01\rangle + |10\rangle \sim 
    \begin{bmatrix}
        - &XX\\
        + &ZZ
    \end{bmatrix}\\
    D&\propto|00\rangle - |11\rangle \sim 
    \begin{bmatrix}
        + &XX\\
        - &ZZ
    \end{bmatrix}\\
\end{align}

A crucial restriction in our representation is based on the realization that a purification circuit consists of two simple stages: bi-local unitary gates that act as permutations in the Bell basis and a coincidence measurement that acts as post-selection, zeroing out exactly half of the components of the diagonal of the density matrix. The group structure of such permutation circuits has been studied extensively~\cite{dehaene2003local, jansen2022enumerating}. A good purification circuit is one for which the Bell basis permutation that it performs is the one that maximizes the probability to be left with the desired final state.
Insight from that group structure has been used to enumerate good purification circuits~\cite{jansen2022enumerating}.

These restrictions are not limiting us when studying entanglement purification: noisy entangled states with a density matrix non-diagonal in the Bell basis can easily be converted into diagonal ones through a modified twirling operation without changing the output fidelity of the purification circuits. Also, Bell-diagonal states naturally arise with realistic noise models such as dephasing and depolarizing.

\subsection{More efficient state representation\label{sec: state rep}} 

Each gate in the purification circuit, and thus the whole circuit, can be thought of as a permutation of Bell basis, and measurement can be thought of as cutting the space of possibilities in two. Such operations preserve the block-diagonal form of the tableaux and only change the phases column. Thus we only need to track these phases. Consider the following state 

\begin{equation}
    A\otimes B\otimes D \sim
    \begin{bmatrix}
        + &XX & &\\
        + &ZZ & &\\
        - & & XX & \\
        - & & ZZ & \\
        + & & &XX\\
        - & & &ZZ
    \end{bmatrix},
\end{equation}
where the omitted off-diagonal entries are identities $II$. 

We track only phases of this state, where $+$ is represented by $1$ and $-$ by $0$. Then, the above state can be represented by $110010$. The new representation will only need $2n$ bits to represent a state of $n$ Bell pairs. We call this new representation a diagonal representation. The space complexities of diagonal, state-vector (a.k.a ket or wavefunction), and tableaux representations are summarized in Table~\ref{tab:complexityTable}. Note that state vector representation is fully general, while tableaux represent a subset of possible states, and the diagonal representation covers an even smaller subset, thus trading off generality for efficiency.  

A bi-local gate from a purification circuit is now simply a permutation over the set of integers (represented by the aforementioned bitstrings). As we discuss below, not all such permutations are valid gates.

\begin{table*}
\centering
\begin{tabular}{ | c | c | c | c |}
\hline
formalism: & state vector & tableaux & Bell diagonal \\
\hline
 $\underbrace{A\otimes B\dots}_{n\text{ pairs}} $ & $ \left.\begin{pmatrix}0\\1\\0\\\vdots\\0\end{pmatrix}\right\}\scriptsize \mathcal{O}(2^{2n}) $ & $ \left.\begin{bmatrix}+&XX&\\+&ZZ&\\&&\ddots\end{bmatrix}\right\}\scriptsize \mathcal{O}(2n\times 2n) $ & $ \underbrace{0011\dots}_{2n \text{ bits}} $ \\
 \hline
 space complexity: & exponential & quadratic & linear\\
 \hline
 can represents: & all possible states & stabilizer states & Bell-diagonal states \\
 \hline
\end{tabular}
\caption{\textbf{Comparison between complexities of ket, tableau, and Bell-diagonal representation}. Let $n$ be the number of Bell pairs in the network. We need exponential space to store a complete state vector representation but need only quadratic space for the tableau, and only linear space if we know the state is a Bell state. Importantly, purification circuits map Bell states to Bell states, thus we can use the most efficient description while being able to simulate any Clifford purification protocol.}
 \label{tab:complexityTable}
\end{table*}

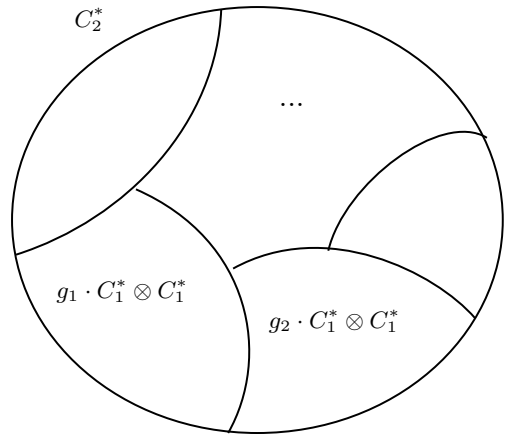
\begin{figure}[h]

\tikzset{every picture/.style={line width=0.75pt}} 

\begin{tikzpicture}[x=0.75pt,y=0.75pt,yscale=-1,xscale=1]

\draw   (256,145.5) .. controls (256,86.13) and (311.4,38) .. (379.75,38) .. controls (448.1,38) and (503.5,86.13) .. (503.5,145.5) .. controls (503.5,204.87) and (448.1,253) .. (379.75,253) .. controls (311.4,253) and (256,204.87) .. (256,145.5) -- cycle ;
\draw    (258,163) .. controls (314.5,145) and (359.5,96) .. (361.5,39) ;
\draw    (365,253) .. controls (386.5,215) and (377.5,156) .. (318.5,130) ;
\draw    (489.5,195) .. controls (469.5,172) and (414.5,144) .. (367.5,170) ;
\draw    (495.5,104) .. controls (472.5,90) and (423.5,127) .. (415.5,161) ;

\draw (389,85) node [anchor=north west][inner sep=0.75pt]  [font=\large] [align=left] {...};
\draw (277,174) node [anchor=north west][inner sep=0.75pt]   [align=left] {$\displaystyle g_{1} \cdot C_{1}^{*} \otimes C_{1}^{*}$};
\draw (384,189) node [anchor=north west][inner sep=0.75pt]   [align=left] {$\displaystyle g_{2} \cdot C_{1}^{*} \otimes C_{1}^{*}$};
\draw (286,37) node [anchor=north west][inner sep=0.75pt]   [align=left] {$\displaystyle C_{2}^{*}$};

\end{tikzpicture}

\caption{\label{fig:cosets c2} \textbf{Graphical illustration of the decomposition $\mathcal C_2^*/\mathcal C_1^*\otimes \mathcal C_1^*$}. All "phaseless" two-qubit Clifford gates can be decomposed in the product of one of 20 "inherently two-qubit" gates, and two single-qubit phaseless Clifford gates (six options for each). This exposes the coset structure of $\mathcal C_2$.}

\end{figure}


Note that noisy dynamics still require a Monte Carlo simulation, but unlike the Clifford approach, we only need $\mathcal{O}(1)$ complexity per gate instead of $\mathcal{O}(N)$. This improvement in efficiency can be thought of as hard-coding the sparse nature of the tableau. The new representation is able to model any Pauli noise.

Our diagonal representation also has a similar performance to the "Pauli frames" approach, but our diagonal representation does not require the simulation of reference frames. Moreover, the more constrained representation of permitted Bell-purifying gates was crucial for the efficient search for good purification circuits -- a feature lacking in Pauli frame simulations.

\subsection{"Bell Preserving" gates}

Above we discussed the state representation. Here we follow with a discussion of the dynamics of these states. We consider only operations that preserve the Bell-diagonal nature of our states. Let us call the set of such gates "Bell Permuting" or "Bell Preserving" (BP). Let us denote the set of BP gates on $n$ Bell-pairs as $\mathcal{B}_n$. It has been known for a while what that group is, and it even has its own (as efficient as Clifford) implementations based on symplectic matrices~\cite{dehaene2003local}. Our work can be restated as the simple realization that these operations have much more compact representation as (canonically enumerated) permutations. We further describe a simple "fixed length" representation (and enumeration) of these BP gates, particularly convenient for software simulations or random sampling.

First note that BP gates are a subset of Clifford gates. Denote all Clifford gates on $N$ qubits as $\mathcal{C}_N$ (for which we also can derive $ |\mathcal{C}_N| = 2(4^N-1)4^N|\mathcal{C}_{N-1}|$). We also use $\mathcal{C}^*_N$ to denote Clifford gates that do not specify phases, i.e., $\mathcal{C}^*_N = \mathcal{C}_N/\mathcal{P}_N$ where $\mathcal{P}_N=\{\sigma_1\otimes\dots\otimes\sigma_N|\sigma_i\in \{I,X,Y,Z\}\}$. Thus we have $\mathcal{C}_N = \left\{C\cdot P\ |\ P\in\mathcal{P}_N , C\in\mathcal{C}^*_N\right\}$ and $ |\mathcal{C}^*_N| = \frac{1}{4^N}|\mathcal{C}_N|$.
              
Observe that Bell Preserving gates on n Bell pairs are always bi-local Clifford gates. Similarly to the introduction of Clifford gates up-to-a-phase, we introduce
  $$ \mathcal{B}^*_n = \mathcal{B}_n / \mathcal{P}_n,$$ for which there is a known identification~\cite{dehaene2003local}  $$ \mathcal{B}^*_n = \mathcal{B}_n / \mathcal{P}_n = \mathrm{Sp}(2n, \mathbb{F}_2).$$
  
We focus on the group $\B_2$ as that is the set of operations typically available in a quantum computer (i.e., two-qubit gates). We can observe that $\{\underbrace{g}_{\tiny Alice}\otimes \underbrace{g}_{\tiny Bob}\ |\ g\in\mathcal{C}^*_2\} \subset \mathcal{B}^*_2$ and that they are of equal size. This simple counting argument gives us $$ \mathcal{B}^*_2 = \mathrm{Sp}(4, \mathbb{F}_2) = \{\underbrace{g}_{\tiny Alice}\otimes \underbrace{g}_{\tiny Bob}\ |\ g\in\mathcal{C}^*_2\}.$$ The $n>2$ case is briefly discussed in the appendix, but the counting argument fails for it and there is no such simple representation.

Lastly, to complete the enumeration, and to provide efficient sampling, we need a convenient way to represent $\mathcal{C}^*_2$. Fig.~\ref{fig:cosets c2} depicts a useful decomposition in a set of cosets $ Q = \mathcal{C}^*_2 / (\mathcal{C}^*_1 \otimes \mathcal{C}^*_1) $, which provides 20 distinct "inherently two-qubit" gates (i.e., $|Q|=20$). This coset decomposition lets us separate the "inherently two-qubit" part of the circuit from the single-qubit dynamics. However, there is no one unique way to pick an element from each coset, so we are left with some arbitrariness in the choice of the 20 gates. We pick one particular set for our software implementation, such that gates like SWAP and CNOT are contained in it for convenience, and with slight abuse of notation we also denote it $Q$.

\begin{figure}
\includegraphics[width = \linewidth]{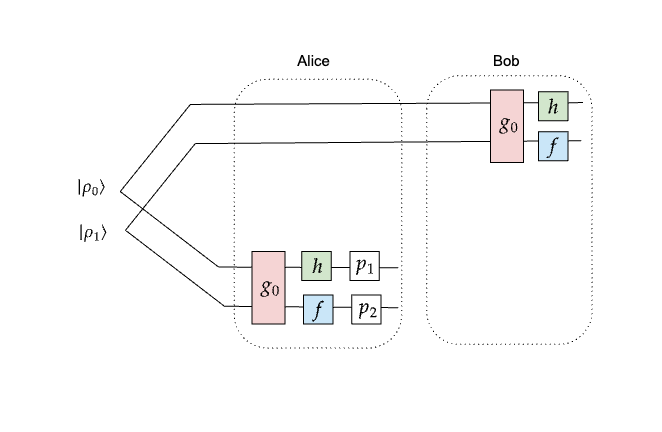}
\caption{\label{fig:twocircuit} \textbf{Circuit representation of a general Bell-Preserving gate using the generators in a compact fixed-length parameterization}. Except for $p_1, p_2 \in \mathcal{P}_1 $ Pauli gates (4 possibilities each), the circuit is symmetric between Alice and Bob. Namely, they both perform $h, f\in\mathcal{C}^*_1$ single-qubit Cliffords (6 possibilities each), and  $g_0 \in Q = \mathcal{C}^*_2 / (\mathcal{C}^*_1\otimes\mathcal{C}^*_1) $ "inherently multiqubit" gates ($20$ possibilities).}
\end{figure}

Putting all this together, a Bell-preserving $\mathcal{B}_2$ gate can be written as \[\underbrace{(g.p)}_{\tiny Alice}\otimes\underbrace{(g)}_{\tiny Bob}\] where $p\in\mathcal{P}_2$ and $g\in\mathcal{C}^*_2$ can be written as \[g = g_0.(h\otimes f),\]
where $h\in\mathcal{C}^*_1$ and $f\in\mathcal{C}^*_1$ and $g_0\in Q$. Thus, a general gate in $\B_2$ can be represented as  in Fig.~\ref{fig:twocircuit}, where $p_1, p_2 \in \mathcal{P}_1 $ Pauli gates (4 possibilities each), $h, f\in\mathcal{C}^*_1$ single-qubit Cliffords (6 possibilities each), and  $g_0 \in Q = \mathcal{C}^*_2 / (\mathcal{C}^*_1\otimes\mathcal{C}^*_1) $ "inherently multiqubit" gates ($20$ possibilities). Thus,  we have \[|\B_2| = 20\times6\times6\times4\times4 = 720\times16 = 11520\] different 2-pairs BP gates. The coset decomposition is partially discussed in~\cite{koenig2014efficiently,jansen2022enumerating,goodenough2023near}.

Lastly, it is useful to restrict oneself to an even smaller set of gates, namely gates that act as the identity when restricted to "good" entangled states, i.e., permutations for which $|AA\rangle\mapsto|AA\rangle$. These "good" gates are a subset of the aforementioned Bell permuting gates and we have provided an interface for it in our implementation of the simulator.

\subsection{Software implementation}

One immediate application of the above result is a more efficient implementation of a simulator for purification circuits. Specifically, instead of working with stabilizer tableaux and arbitrary bi-local four-qubit gates, we can decompose each of the BP gates into one of the 20 two-qubit gates ($Q$), two of the 6 one-qubit phase-less gates ($C_1^*$), and two Pauli gates. Internally, each of these subgroups of gates is represented as a subgroup of permutations over integers and the bitstring representation of the integers coincides with a particular set of Bell pairs as discussed above. Thus any bi-lateral gate can be fully defined by 5 indices. With the diagonal state representation, the application of any of the 11520 BP gates takes $\mathcal{O}(1)$ steps (simply a mapping from one bitstring (a 2-bit or 4-bit integer) to another bitstring. 

Using diagonal state representation and Bell-Preserving gates, we can significantly improve the performance of purification circuit optimization since the gate representation is extremely compact and of fixed length and it takes constant time to apply. 

\begin{figure}
    \centering
    \includegraphics[width =\linewidth]{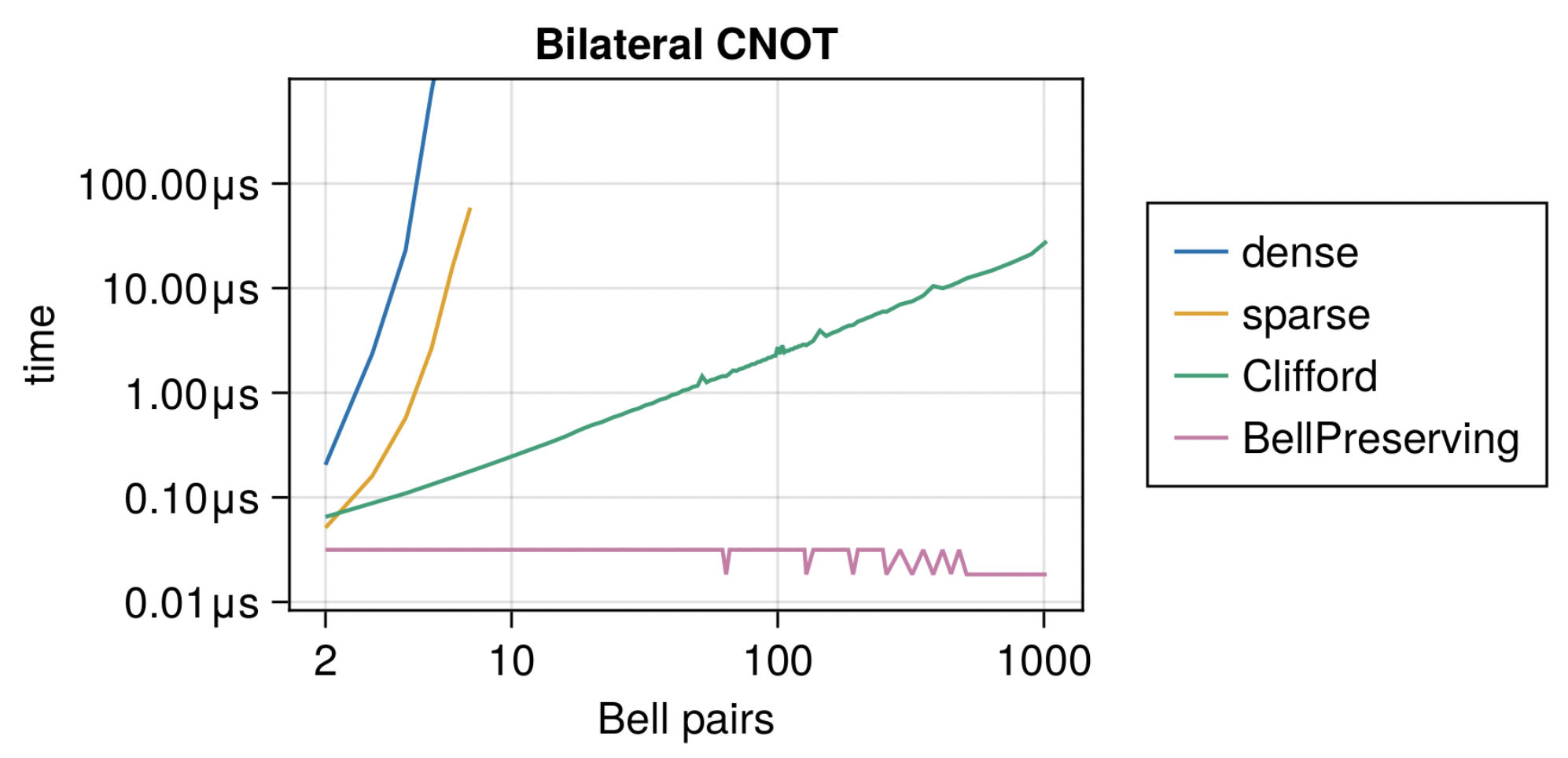}
    \caption{\textbf{Comparison between run time using ket, tableau, and Bell-diagonal representations}. As expected, a full-state vector simulation is exponentially expensive, independently of whether the operator is represented as a sparse or a dense matrix. Much more interesting is the comparison between a typical Clifford circuit simulation (here using the highly optimized QuantumClifford.jl implementation) and our new BPGates.jl implementation based on the methods described in this paper. Additionally, BPGates.jl provides for very compact parameterization of all possible purification circuit gates, lending itself to use in circuit optimizers.}
    \label{fig:complexityplot}
\end{figure}

The BP representation is implemented in Julia as the BPGate.jl package. We compare the performance against a thoroughly optimized tableaux simulator (QuantumClifford.jl) and wavefunction simulator (QuantumOptics.jl, including both dense and sparse matrix representations of operators). In Fig.~\ref{fig:complexityplot}, we plot the run time of applying a bilateral CNOT gate to two randomly chosen pairs of Bell pairs in the network against the number of Bell pairs in the entanglement network. The run time of BPGates is independent of the number of Bell pairs in the network ($\mathcal{O}(1)$), which improves from the linear runtime of the Clifford representation and exponential runtime using the matrix representation.

Above we have discussed the implementation of $\B_2$, but a purification circuit may involve permutation of more than 2 Bell pairs. Naturally, any such larger circuit would need to be decomposed in two-qubit gates to be executed on a quantum device. We have also restricted ourselves only to bi-partite entanglement, but similar "block-diagonal tableaux" and "base-preserving gate" formalism exist for the purification of any multipartite entangled stabilizer state.

\section{Discussion and Conclusion}

This paper introduces three important new directions to the theory and practice of quantum networking. The new entanglement purification simulation algorithm we introduce has much lower complexity than the alternatives. Moreover, it provides for very natural parameterization of permitted gates in a purification circuit. Of course, with higher performance, comes specialization, thus the algorithm is applicable only to distillation circuits.

This algorithm enabled circuit optimization at a previously impossible scale. Not only was the cost of simulating a single state too high, but due to the unstructured nature of the search space a vast array of simulations need to be executed. This is now practical and any new search method can benefit from the improvements we have enabled. The optimized circuits we can now generate are superior to anything else available in the literature and can be fine-tuned to the particular noise model of the hardware that will be executing them.

Lastly, we demonstrate that choices of figures of merit used by virtually all literature on the topic can be misguided. In the typically studied asymptotic case, the difference is insignificant, but when one considers the constraints of a finite-size quantum register, it becomes important to employ a specialized figure of merit. For instance, we show that if the entanglement purification is used as the physical layer under the teleportation of logical qubits, then the typical definition of fidelity (denoted $F_A$ here) is the one that leads to the worst possible performance. This happens because $F_A$ does not distinguish between low-weight and high-weight correlated errors, while this distinction is crucial to the performance of the error correction code at the next layer of the technology stack.

There are numerous further steps that can be taken to improve these results. Our superior simulation algorithm can be repurposed for the modeling of the purification of arbitrary graph states. That would require generating a new set of "good" gates (and corresponding permutations) for each class of graph states. Thankfully this process can in principle be automated in software. But even restricting ourselves to cases of Bell pair purification there is much more that can be done on the engineering side, including more detailed parameter sweeps of best circuits for various types of hardware, and for various applications (similarly to the entanglement teleportation example we gave in this work). Lastly, there is a rich design space to be explored that studies how to best create high-fidelity multi-party entangled states -- in what order should we nest Bell state generation, Bell state purification, entanglement swapping, merging of Bell pairs into larger graph states, and graph state purification itself. Many of these tasks are now much easier to model thanks to our work, but there is still much more to explore.

\paragraph*{Code availability:} The underlying simulator is available as a registered package in the Julia ecosystem: BPGates.jl~\cite{stefan_krastanov_2023_8137209}. The optimization routines that we used on top of that simulator are available as easy to reproduce scripts at~\cite{vaishnavi_addala_2023_7844179}.

\begin{acknowledgements}
V.A. performed the design and optimization of entanglement circuits. S.G. developed the new fast simulation methods. S.K. designed and directed the overall project. The manuscript was prepared by the authors together.

We thank Isaac Chuang, Annie Wei, and Dirk Englund for their helpful comments and support. The MIT Supercloud, MIT UROP office, and the Reed Fund provided invaluable resources. This project would not have been possible without the contributions of the Julia open source projects. This work was supported by the Center for Quantum Networks, grant ERC 1941583.
\end{acknowledgements}

\bibliography{bibliography}

\begin{thebibliography}{38}%
\makeatletter
\providecommand \@ifxundefined [1]{%
 \@ifx{#1\undefined}
}%
\providecommand \@ifnum [1]{%
 \ifnum #1\expandafter \@firstoftwo
 \else \expandafter \@secondoftwo
 \fi
}%
\providecommand \@ifx [1]{%
 \ifx #1\expandafter \@firstoftwo
 \else \expandafter \@secondoftwo
 \fi
}%
\providecommand \natexlab [1]{#1}%
\providecommand \enquote  [1]{``#1''}%
\providecommand \bibnamefont  [1]{#1}%
\providecommand \bibfnamefont [1]{#1}%
\providecommand \citenamefont [1]{#1}%
\providecommand \href@noop [0]{\@secondoftwo}%
\providecommand \href [0]{\begingroup \@sanitize@url \@href}%
\providecommand \@href[1]{\@@startlink{#1}\@@href}%
\providecommand \@@href[1]{\endgroup#1\@@endlink}%
\providecommand \@sanitize@url [0]{\catcode `\\12\catcode `\$12\catcode
  `\&12\catcode `\#12\catcode `\^12\catcode `\_12\catcode `\%12\relax}%
\providecommand \@@startlink[1]{}%
\providecommand \@@endlink[0]{}%
\providecommand \url  [0]{\begingroup\@sanitize@url \@url }%
\providecommand \@url [1]{\endgroup\@href {#1}{\urlprefix }}%
\providecommand \urlprefix  [0]{URL }%
\providecommand \Eprint [0]{\href }%
\providecommand \doibase [0]{https://doi.org/}%
\providecommand \selectlanguage [0]{\@gobble}%
\providecommand \bibinfo  [0]{\@secondoftwo}%
\providecommand \bibfield  [0]{\@secondoftwo}%
\providecommand \translation [1]{[#1]}%
\providecommand \BibitemOpen [0]{}%
\providecommand \bibitemStop [0]{}%
\providecommand \bibitemNoStop [0]{.\EOS\space}%
\providecommand \EOS [0]{\spacefactor3000\relax}%
\providecommand \BibitemShut  [1]{\csname bibitem#1\endcsname}%
\let\auto@bib@innerbib\@empty
\bibitem [{\citenamefont {Hucul}\ \emph {et~al.}(2015)\citenamefont {Hucul},
  \citenamefont {Inlek}, \citenamefont {Vittorini}, \citenamefont {Crocker},
  \citenamefont {Debnath}, \citenamefont {Clark},\ and\ \citenamefont
  {Monroe}}]{hucul2015modular}%
  \BibitemOpen
  \bibfield  {author} {\bibinfo {author} {\bibfnamefont {D.}~\bibnamefont
  {Hucul}}, \bibinfo {author} {\bibfnamefont {I.}~\bibnamefont {Inlek}},
  \bibinfo {author} {\bibfnamefont {G.}~\bibnamefont {Vittorini}}, \bibinfo
  {author} {\bibfnamefont {C.}~\bibnamefont {Crocker}}, \bibinfo {author}
  {\bibfnamefont {S.}~\bibnamefont {Debnath}}, \bibinfo {author} {\bibfnamefont
  {S.}~\bibnamefont {Clark}},\ and\ \bibinfo {author} {\bibfnamefont
  {C.}~\bibnamefont {Monroe}},\ }\href@noop {} {\bibfield  {journal} {\bibinfo
  {journal} {Nature Physics}\ }\textbf {\bibinfo {volume} {11}},\ \bibinfo
  {pages} {37} (\bibinfo {year} {2015})}\BibitemShut {NoStop}%
\bibitem [{\citenamefont {Moehring}\ \emph {et~al.}(2007)\citenamefont
  {Moehring}, \citenamefont {Maunz}, \citenamefont {Olmschenk}, \citenamefont
  {Younge}, \citenamefont {Matsukevich}, \citenamefont {Duan},\ and\
  \citenamefont {Monroe}}]{moehring2007entanglement}%
  \BibitemOpen
  \bibfield  {author} {\bibinfo {author} {\bibfnamefont {D.}~\bibnamefont
  {Moehring}}, \bibinfo {author} {\bibfnamefont {P.}~\bibnamefont {Maunz}},
  \bibinfo {author} {\bibfnamefont {S.}~\bibnamefont {Olmschenk}}, \bibinfo
  {author} {\bibfnamefont {K.}~\bibnamefont {Younge}}, \bibinfo {author}
  {\bibfnamefont {D.}~\bibnamefont {Matsukevich}}, \bibinfo {author}
  {\bibfnamefont {L.-M.}\ \bibnamefont {Duan}},\ and\ \bibinfo {author}
  {\bibfnamefont {C.}~\bibnamefont {Monroe}},\ }\href@noop {} {\bibfield
  {journal} {\bibinfo  {journal} {Nature}\ }\textbf {\bibinfo {volume} {449}},\
  \bibinfo {pages} {68} (\bibinfo {year} {2007})}\BibitemShut {NoStop}%
\bibitem [{\citenamefont {Krutyanskiy}\ \emph {et~al.}(2023)\citenamefont
  {Krutyanskiy}, \citenamefont {Galli}, \citenamefont {Krcmarsky},
  \citenamefont {Baier}, \citenamefont {Fioretto}, \citenamefont {Pu},
  \citenamefont {Mazloom}, \citenamefont {Sekatski}, \citenamefont {Canteri},
  \citenamefont {Teller} \emph {et~al.}}]{krutyanskiy2023entanglement}%
  \BibitemOpen
  \bibfield  {author} {\bibinfo {author} {\bibfnamefont {V.}~\bibnamefont
  {Krutyanskiy}}, \bibinfo {author} {\bibfnamefont {M.}~\bibnamefont {Galli}},
  \bibinfo {author} {\bibfnamefont {V.}~\bibnamefont {Krcmarsky}}, \bibinfo
  {author} {\bibfnamefont {S.}~\bibnamefont {Baier}}, \bibinfo {author}
  {\bibfnamefont {D.}~\bibnamefont {Fioretto}}, \bibinfo {author}
  {\bibfnamefont {Y.}~\bibnamefont {Pu}}, \bibinfo {author} {\bibfnamefont
  {A.}~\bibnamefont {Mazloom}}, \bibinfo {author} {\bibfnamefont
  {P.}~\bibnamefont {Sekatski}}, \bibinfo {author} {\bibfnamefont
  {M.}~\bibnamefont {Canteri}}, \bibinfo {author} {\bibfnamefont
  {M.}~\bibnamefont {Teller}}, \emph {et~al.},\ }\href@noop {} {\bibfield
  {journal} {\bibinfo  {journal} {Physical Review Letters}\ }\textbf {\bibinfo
  {volume} {130}},\ \bibinfo {pages} {050803} (\bibinfo {year}
  {2023})}\BibitemShut {NoStop}%
\bibitem [{\citenamefont {Pfaff}\ \emph {et~al.}(2013)\citenamefont {Pfaff},
  \citenamefont {Taminiau}, \citenamefont {Robledo}, \citenamefont {Bernien},
  \citenamefont {Markham}, \citenamefont {Twitchen},\ and\ \citenamefont
  {Hanson}}]{pfaff2013demonstration}%
  \BibitemOpen
  \bibfield  {author} {\bibinfo {author} {\bibfnamefont {W.}~\bibnamefont
  {Pfaff}}, \bibinfo {author} {\bibfnamefont {T.~H.}\ \bibnamefont {Taminiau}},
  \bibinfo {author} {\bibfnamefont {L.}~\bibnamefont {Robledo}}, \bibinfo
  {author} {\bibfnamefont {H.}~\bibnamefont {Bernien}}, \bibinfo {author}
  {\bibfnamefont {M.}~\bibnamefont {Markham}}, \bibinfo {author} {\bibfnamefont
  {D.~J.}\ \bibnamefont {Twitchen}},\ and\ \bibinfo {author} {\bibfnamefont
  {R.}~\bibnamefont {Hanson}},\ }\href@noop {} {\bibfield  {journal} {\bibinfo
  {journal} {Nature Physics}\ }\textbf {\bibinfo {volume} {9}},\ \bibinfo
  {pages} {29} (\bibinfo {year} {2013})}\BibitemShut {NoStop}%
\bibitem [{\citenamefont {Hensen}\ \emph {et~al.}(2015)\citenamefont {Hensen},
  \citenamefont {Bernien}, \citenamefont {Dr{\'e}au}, \citenamefont {Reiserer},
  \citenamefont {Kalb}, \citenamefont {Blok}, \citenamefont {Ruitenberg},
  \citenamefont {Vermeulen}, \citenamefont {Schouten}, \citenamefont
  {Abell{\'a}n} \emph {et~al.}}]{hensen2015loophole}%
  \BibitemOpen
  \bibfield  {author} {\bibinfo {author} {\bibfnamefont {B.}~\bibnamefont
  {Hensen}}, \bibinfo {author} {\bibfnamefont {H.}~\bibnamefont {Bernien}},
  \bibinfo {author} {\bibfnamefont {A.~E.}\ \bibnamefont {Dr{\'e}au}}, \bibinfo
  {author} {\bibfnamefont {A.}~\bibnamefont {Reiserer}}, \bibinfo {author}
  {\bibfnamefont {N.}~\bibnamefont {Kalb}}, \bibinfo {author} {\bibfnamefont
  {M.~S.}\ \bibnamefont {Blok}}, \bibinfo {author} {\bibfnamefont
  {J.}~\bibnamefont {Ruitenberg}}, \bibinfo {author} {\bibfnamefont {R.~F.}\
  \bibnamefont {Vermeulen}}, \bibinfo {author} {\bibfnamefont {R.~N.}\
  \bibnamefont {Schouten}}, \bibinfo {author} {\bibfnamefont {C.}~\bibnamefont
  {Abell{\'a}n}}, \emph {et~al.},\ }\href@noop {} {\bibfield  {journal}
  {\bibinfo  {journal} {Nature}\ }\textbf {\bibinfo {volume} {526}},\ \bibinfo
  {pages} {682} (\bibinfo {year} {2015})}\BibitemShut {NoStop}%
\bibitem [{\citenamefont {Hermans}\ \emph {et~al.}(2023)\citenamefont
  {Hermans}, \citenamefont {Pompili}, \citenamefont {Martins}, \citenamefont
  {Montblanch}, \citenamefont {Beukers}, \citenamefont {Baier}, \citenamefont
  {Borregaard},\ and\ \citenamefont {Hanson}}]{hermans2023entangling}%
  \BibitemOpen
  \bibfield  {author} {\bibinfo {author} {\bibfnamefont {S.~L.~N.}\
  \bibnamefont {Hermans}}, \bibinfo {author} {\bibfnamefont {M.}~\bibnamefont
  {Pompili}}, \bibinfo {author} {\bibfnamefont {L.~D.~S.}\ \bibnamefont
  {Martins}}, \bibinfo {author} {\bibfnamefont {A.~R.-P.}\ \bibnamefont
  {Montblanch}}, \bibinfo {author} {\bibfnamefont {H.~K.~C.}\ \bibnamefont
  {Beukers}}, \bibinfo {author} {\bibfnamefont {S.}~\bibnamefont {Baier}},
  \bibinfo {author} {\bibfnamefont {J.}~\bibnamefont {Borregaard}},\ and\
  \bibinfo {author} {\bibfnamefont {R.}~\bibnamefont {Hanson}},\ }\href@noop {}
  {\bibfield  {journal} {\bibinfo  {journal} {New Journal of Physics}\ }\textbf
  {\bibinfo {volume} {25}},\ \bibinfo {pages} {013011} (\bibinfo {year}
  {2023})}\BibitemShut {NoStop}%
\bibitem [{\citenamefont {Ritter}\ \emph {et~al.}(2012)\citenamefont {Ritter},
  \citenamefont {N{\"o}lleke}, \citenamefont {Hahn}, \citenamefont {Reiserer},
  \citenamefont {Neuzner}, \citenamefont {Uphoff}, \citenamefont {M{\"u}cke},
  \citenamefont {Figueroa}, \citenamefont {Bochmann},\ and\ \citenamefont
  {Rempe}}]{ritter2012elementary}%
  \BibitemOpen
  \bibfield  {author} {\bibinfo {author} {\bibfnamefont {S.}~\bibnamefont
  {Ritter}}, \bibinfo {author} {\bibfnamefont {C.}~\bibnamefont {N{\"o}lleke}},
  \bibinfo {author} {\bibfnamefont {C.}~\bibnamefont {Hahn}}, \bibinfo {author}
  {\bibfnamefont {A.}~\bibnamefont {Reiserer}}, \bibinfo {author}
  {\bibfnamefont {A.}~\bibnamefont {Neuzner}}, \bibinfo {author} {\bibfnamefont
  {M.}~\bibnamefont {Uphoff}}, \bibinfo {author} {\bibfnamefont
  {M.}~\bibnamefont {M{\"u}cke}}, \bibinfo {author} {\bibfnamefont
  {E.}~\bibnamefont {Figueroa}}, \bibinfo {author} {\bibfnamefont
  {J.}~\bibnamefont {Bochmann}},\ and\ \bibinfo {author} {\bibfnamefont
  {G.}~\bibnamefont {Rempe}},\ }\href@noop {} {\bibfield  {journal} {\bibinfo
  {journal} {Nature}\ }\textbf {\bibinfo {volume} {484}},\ \bibinfo {pages}
  {195} (\bibinfo {year} {2012})}\BibitemShut {NoStop}%
\bibitem [{\citenamefont {Yang}\ \emph {et~al.}(2022)\citenamefont {Yang},
  \citenamefont {Yu}, \citenamefont {Li}, \citenamefont {Jing}, \citenamefont
  {Bao},\ and\ \citenamefont {Pan}}]{yang2022sequential}%
  \BibitemOpen
  \bibfield  {author} {\bibinfo {author} {\bibfnamefont {C.-W.}\ \bibnamefont
  {Yang}}, \bibinfo {author} {\bibfnamefont {Y.}~\bibnamefont {Yu}}, \bibinfo
  {author} {\bibfnamefont {J.}~\bibnamefont {Li}}, \bibinfo {author}
  {\bibfnamefont {B.}~\bibnamefont {Jing}}, \bibinfo {author} {\bibfnamefont
  {X.-H.}\ \bibnamefont {Bao}},\ and\ \bibinfo {author} {\bibfnamefont {J.-W.}\
  \bibnamefont {Pan}},\ }\href@noop {} {\bibfield  {journal} {\bibinfo
  {journal} {Nature Photonics}\ }\textbf {\bibinfo {volume} {16}},\ \bibinfo
  {pages} {658} (\bibinfo {year} {2022})}\BibitemShut {NoStop}%
\bibitem [{\citenamefont {van Leent}\ \emph {et~al.}(2022)\citenamefont {van
  Leent}, \citenamefont {Bock}, \citenamefont {Fertig}, \citenamefont
  {Garthoff}, \citenamefont {Eppelt}, \citenamefont {Zhou}, \citenamefont
  {Malik}, \citenamefont {Seubert}, \citenamefont {Bauer}, \citenamefont
  {Rosenfeld} \emph {et~al.}}]{van2022entangling}%
  \BibitemOpen
  \bibfield  {author} {\bibinfo {author} {\bibfnamefont {T.}~\bibnamefont {van
  Leent}}, \bibinfo {author} {\bibfnamefont {M.}~\bibnamefont {Bock}}, \bibinfo
  {author} {\bibfnamefont {F.}~\bibnamefont {Fertig}}, \bibinfo {author}
  {\bibfnamefont {R.}~\bibnamefont {Garthoff}}, \bibinfo {author}
  {\bibfnamefont {S.}~\bibnamefont {Eppelt}}, \bibinfo {author} {\bibfnamefont
  {Y.}~\bibnamefont {Zhou}}, \bibinfo {author} {\bibfnamefont {P.}~\bibnamefont
  {Malik}}, \bibinfo {author} {\bibfnamefont {M.}~\bibnamefont {Seubert}},
  \bibinfo {author} {\bibfnamefont {T.}~\bibnamefont {Bauer}}, \bibinfo
  {author} {\bibfnamefont {W.}~\bibnamefont {Rosenfeld}}, \emph {et~al.},\
  }\href@noop {} {\bibfield  {journal} {\bibinfo  {journal} {Nature}\ }\textbf
  {\bibinfo {volume} {607}},\ \bibinfo {pages} {69} (\bibinfo {year}
  {2022})}\BibitemShut {NoStop}%
\bibitem [{\citenamefont {Yu}\ \emph {et~al.}(2020)\citenamefont {Yu},
  \citenamefont {Ma}, \citenamefont {Luo}, \citenamefont {Jing}, \citenamefont
  {Sun}, \citenamefont {Fang}, \citenamefont {Yang}, \citenamefont {Liu},
  \citenamefont {Zheng}, \citenamefont {Xie} \emph
  {et~al.}}]{yu2020entanglement}%
  \BibitemOpen
  \bibfield  {author} {\bibinfo {author} {\bibfnamefont {Y.}~\bibnamefont
  {Yu}}, \bibinfo {author} {\bibfnamefont {F.}~\bibnamefont {Ma}}, \bibinfo
  {author} {\bibfnamefont {X.-Y.}\ \bibnamefont {Luo}}, \bibinfo {author}
  {\bibfnamefont {B.}~\bibnamefont {Jing}}, \bibinfo {author} {\bibfnamefont
  {P.-F.}\ \bibnamefont {Sun}}, \bibinfo {author} {\bibfnamefont {R.-Z.}\
  \bibnamefont {Fang}}, \bibinfo {author} {\bibfnamefont {C.-W.}\ \bibnamefont
  {Yang}}, \bibinfo {author} {\bibfnamefont {H.}~\bibnamefont {Liu}}, \bibinfo
  {author} {\bibfnamefont {M.-Y.}\ \bibnamefont {Zheng}}, \bibinfo {author}
  {\bibfnamefont {X.-P.}\ \bibnamefont {Xie}}, \emph {et~al.},\ }\href@noop {}
  {\bibfield  {journal} {\bibinfo  {journal} {Nature}\ }\textbf {\bibinfo
  {volume} {578}},\ \bibinfo {pages} {240} (\bibinfo {year}
  {2020})}\BibitemShut {NoStop}%
\bibitem [{\citenamefont {Luo}\ \emph {et~al.}(2022)\citenamefont {Luo},
  \citenamefont {Yu}, \citenamefont {Liu}, \citenamefont {Zheng}, \citenamefont
  {Wang}, \citenamefont {Wang}, \citenamefont {Li}, \citenamefont {Jiang},
  \citenamefont {Xie}, \citenamefont {Zhang} \emph
  {et~al.}}]{luo2022postselected}%
  \BibitemOpen
  \bibfield  {author} {\bibinfo {author} {\bibfnamefont {X.-Y.}\ \bibnamefont
  {Luo}}, \bibinfo {author} {\bibfnamefont {Y.}~\bibnamefont {Yu}}, \bibinfo
  {author} {\bibfnamefont {J.-L.}\ \bibnamefont {Liu}}, \bibinfo {author}
  {\bibfnamefont {M.-Y.}\ \bibnamefont {Zheng}}, \bibinfo {author}
  {\bibfnamefont {C.-Y.}\ \bibnamefont {Wang}}, \bibinfo {author}
  {\bibfnamefont {B.}~\bibnamefont {Wang}}, \bibinfo {author} {\bibfnamefont
  {J.}~\bibnamefont {Li}}, \bibinfo {author} {\bibfnamefont {X.}~\bibnamefont
  {Jiang}}, \bibinfo {author} {\bibfnamefont {X.-P.}\ \bibnamefont {Xie}},
  \bibinfo {author} {\bibfnamefont {Q.}~\bibnamefont {Zhang}}, \emph {et~al.},\
  }\href@noop {} {\bibfield  {journal} {\bibinfo  {journal} {Physical Review
  Letters}\ }\textbf {\bibinfo {volume} {129}},\ \bibinfo {pages} {050503}
  (\bibinfo {year} {2022})}\BibitemShut {NoStop}%
\bibitem [{\citenamefont {Narla}\ \emph {et~al.}(2016)\citenamefont {Narla},
  \citenamefont {Shankar}, \citenamefont {Hatridge}, \citenamefont {Leghtas},
  \citenamefont {Sliwa}, \citenamefont {Zalys-Geller}, \citenamefont
  {Mundhada}, \citenamefont {Pfaff}, \citenamefont {Frunzio}, \citenamefont
  {Schoelkopf} \emph {et~al.}}]{narla2016robust}%
  \BibitemOpen
  \bibfield  {author} {\bibinfo {author} {\bibfnamefont {A.}~\bibnamefont
  {Narla}}, \bibinfo {author} {\bibfnamefont {S.}~\bibnamefont {Shankar}},
  \bibinfo {author} {\bibfnamefont {M.}~\bibnamefont {Hatridge}}, \bibinfo
  {author} {\bibfnamefont {Z.}~\bibnamefont {Leghtas}}, \bibinfo {author}
  {\bibfnamefont {K.}~\bibnamefont {Sliwa}}, \bibinfo {author} {\bibfnamefont
  {E.}~\bibnamefont {Zalys-Geller}}, \bibinfo {author} {\bibfnamefont
  {S.}~\bibnamefont {Mundhada}}, \bibinfo {author} {\bibfnamefont
  {W.}~\bibnamefont {Pfaff}}, \bibinfo {author} {\bibfnamefont
  {L.}~\bibnamefont {Frunzio}}, \bibinfo {author} {\bibfnamefont
  {R.}~\bibnamefont {Schoelkopf}}, \emph {et~al.},\ }\href@noop {} {\bibfield
  {journal} {\bibinfo  {journal} {Physical Review X}\ }\textbf {\bibinfo
  {volume} {6}},\ \bibinfo {pages} {031036} (\bibinfo {year}
  {2016})}\BibitemShut {NoStop}%
\bibitem [{\citenamefont {Rakonjac}\ \emph {et~al.}(2023)\citenamefont
  {Rakonjac}, \citenamefont {Grandi}, \citenamefont {Wengerowsky},
  \citenamefont {Lago-Rivera}, \citenamefont {Appas},\ and\ \citenamefont
  {de~Riedmatten}}]{rakonjac2023transmission}%
  \BibitemOpen
  \bibfield  {author} {\bibinfo {author} {\bibfnamefont {J.~V.}\ \bibnamefont
  {Rakonjac}}, \bibinfo {author} {\bibfnamefont {S.}~\bibnamefont {Grandi}},
  \bibinfo {author} {\bibfnamefont {S.}~\bibnamefont {Wengerowsky}}, \bibinfo
  {author} {\bibfnamefont {D.}~\bibnamefont {Lago-Rivera}}, \bibinfo {author}
  {\bibfnamefont {F.}~\bibnamefont {Appas}},\ and\ \bibinfo {author}
  {\bibfnamefont {H.}~\bibnamefont {de~Riedmatten}},\ }\href@noop {} {\bibfield
   {journal} {\bibinfo  {journal} {arXiv preprint arXiv:2304.05416}\ }
  (\bibinfo {year} {2023})}\BibitemShut {NoStop}%
\bibitem [{\citenamefont {Jiang}\ \emph {et~al.}(2022)\citenamefont {Jiang},
  \citenamefont {Xue}, \citenamefont {He}, \citenamefont {An}, \citenamefont
  {Zheng}, \citenamefont {Xu}, \citenamefont {Wen}, \citenamefont {Xie},
  \citenamefont {Lu}, \citenamefont {Zhu} \emph {et~al.}}]{jiang2022quantum}%
  \BibitemOpen
  \bibfield  {author} {\bibinfo {author} {\bibfnamefont {M.-H.}\ \bibnamefont
  {Jiang}}, \bibinfo {author} {\bibfnamefont {W.}~\bibnamefont {Xue}}, \bibinfo
  {author} {\bibfnamefont {Q.}~\bibnamefont {He}}, \bibinfo {author}
  {\bibfnamefont {Y.-Y.}\ \bibnamefont {An}}, \bibinfo {author} {\bibfnamefont
  {X.}~\bibnamefont {Zheng}}, \bibinfo {author} {\bibfnamefont {W.-J.}\
  \bibnamefont {Xu}}, \bibinfo {author} {\bibfnamefont {W.}~\bibnamefont
  {Wen}}, \bibinfo {author} {\bibfnamefont {Y.-B.}\ \bibnamefont {Xie}},
  \bibinfo {author} {\bibfnamefont {Y.}~\bibnamefont {Lu}}, \bibinfo {author}
  {\bibfnamefont {S.}~\bibnamefont {Zhu}}, \emph {et~al.},\ }\href@noop {}
  {\bibfield  {journal} {\bibinfo  {journal} {arXiv preprint arXiv:2212.12898}\
  } (\bibinfo {year} {2022})}\BibitemShut {NoStop}%
\bibitem [{\citenamefont {Deutsch}\ \emph {et~al.}(1996)\citenamefont
  {Deutsch}, \citenamefont {Ekert}, \citenamefont {Jozsa}, \citenamefont
  {Macchiavello}, \citenamefont {Popescu},\ and\ \citenamefont
  {Sanpera}}]{deutsch1996quantum}%
  \BibitemOpen
  \bibfield  {author} {\bibinfo {author} {\bibfnamefont {D.}~\bibnamefont
  {Deutsch}}, \bibinfo {author} {\bibfnamefont {A.}~\bibnamefont {Ekert}},
  \bibinfo {author} {\bibfnamefont {R.}~\bibnamefont {Jozsa}}, \bibinfo
  {author} {\bibfnamefont {C.}~\bibnamefont {Macchiavello}}, \bibinfo {author}
  {\bibfnamefont {S.}~\bibnamefont {Popescu}},\ and\ \bibinfo {author}
  {\bibfnamefont {A.}~\bibnamefont {Sanpera}},\ }\href@noop {} {\bibfield
  {journal} {\bibinfo  {journal} {Physical review letters}\ }\textbf {\bibinfo
  {volume} {77}},\ \bibinfo {pages} {2818} (\bibinfo {year}
  {1996})}\BibitemShut {NoStop}%
\bibitem [{\citenamefont {Bennett}\ \emph
  {et~al.}(1996{\natexlab{a}})\citenamefont {Bennett}, \citenamefont
  {Brassard}, \citenamefont {Popescu}, \citenamefont {Schumacher},
  \citenamefont {Smolin},\ and\ \citenamefont
  {Wootters}}]{bennett1996purification}%
  \BibitemOpen
  \bibfield  {author} {\bibinfo {author} {\bibfnamefont {C.~H.}\ \bibnamefont
  {Bennett}}, \bibinfo {author} {\bibfnamefont {G.}~\bibnamefont {Brassard}},
  \bibinfo {author} {\bibfnamefont {S.}~\bibnamefont {Popescu}}, \bibinfo
  {author} {\bibfnamefont {B.}~\bibnamefont {Schumacher}}, \bibinfo {author}
  {\bibfnamefont {J.~A.}\ \bibnamefont {Smolin}},\ and\ \bibinfo {author}
  {\bibfnamefont {W.~K.}\ \bibnamefont {Wootters}},\ }\href@noop {} {\bibfield
  {journal} {\bibinfo  {journal} {Physical review letters}\ }\textbf {\bibinfo
  {volume} {76}},\ \bibinfo {pages} {722} (\bibinfo {year}
  {1996}{\natexlab{a}})}\BibitemShut {NoStop}%
\bibitem [{\citenamefont {D{\"u}r}\ \emph {et~al.}(1999)\citenamefont
  {D{\"u}r}, \citenamefont {Briegel}, \citenamefont {Cirac},\ and\
  \citenamefont {Zoller}}]{dur1999quantum}%
  \BibitemOpen
  \bibfield  {author} {\bibinfo {author} {\bibfnamefont {W.}~\bibnamefont
  {D{\"u}r}}, \bibinfo {author} {\bibfnamefont {H.-J.}\ \bibnamefont
  {Briegel}}, \bibinfo {author} {\bibfnamefont {J.}~\bibnamefont {Cirac}},\
  and\ \bibinfo {author} {\bibfnamefont {P.}~\bibnamefont {Zoller}},\
  }\href@noop {} {\bibfield  {journal} {\bibinfo  {journal} {Physical Review
  A}\ }\textbf {\bibinfo {volume} {59}},\ \bibinfo {pages} {169} (\bibinfo
  {year} {1999})}\BibitemShut {NoStop}%
\bibitem [{\citenamefont {D{\"u}r}\ and\ \citenamefont
  {Briegel}(2007)}]{dur2007entanglement}%
  \BibitemOpen
  \bibfield  {author} {\bibinfo {author} {\bibfnamefont {W.}~\bibnamefont
  {D{\"u}r}}\ and\ \bibinfo {author} {\bibfnamefont {H.~J.}\ \bibnamefont
  {Briegel}},\ }\href@noop {} {\bibfield  {journal} {\bibinfo  {journal}
  {Reports on Progress in Physics}\ }\textbf {\bibinfo {volume} {70}},\
  \bibinfo {pages} {1381} (\bibinfo {year} {2007})}\BibitemShut {NoStop}%
\bibitem [{\citenamefont {Fujii}\ and\ \citenamefont
  {Yamamoto}(2009)}]{fujii2009entanglement}%
  \BibitemOpen
  \bibfield  {author} {\bibinfo {author} {\bibfnamefont {K.}~\bibnamefont
  {Fujii}}\ and\ \bibinfo {author} {\bibfnamefont {K.}~\bibnamefont
  {Yamamoto}},\ }\href@noop {} {\bibfield  {journal} {\bibinfo  {journal}
  {Physical Review A}\ }\textbf {\bibinfo {volume} {80}},\ \bibinfo {pages}
  {042308} (\bibinfo {year} {2009})}\BibitemShut {NoStop}%
\bibitem [{\citenamefont {Nickerson}\ \emph {et~al.}(2013)\citenamefont
  {Nickerson}, \citenamefont {Li},\ and\ \citenamefont
  {Benjamin}}]{nickerson2013topological}%
  \BibitemOpen
  \bibfield  {author} {\bibinfo {author} {\bibfnamefont {N.~H.}\ \bibnamefont
  {Nickerson}}, \bibinfo {author} {\bibfnamefont {Y.}~\bibnamefont {Li}},\ and\
  \bibinfo {author} {\bibfnamefont {S.~C.}\ \bibnamefont {Benjamin}},\
  }\href@noop {} {\bibfield  {journal} {\bibinfo  {journal} {Nature
  communications}\ }\textbf {\bibinfo {volume} {4}},\ \bibinfo {pages} {1756}
  (\bibinfo {year} {2013})}\BibitemShut {NoStop}%
\bibitem [{\citenamefont {Nickerson}\ \emph {et~al.}(2014)\citenamefont
  {Nickerson}, \citenamefont {Fitzsimons},\ and\ \citenamefont
  {Benjamin}}]{nickerson2014freely}%
  \BibitemOpen
  \bibfield  {author} {\bibinfo {author} {\bibfnamefont {N.~H.}\ \bibnamefont
  {Nickerson}}, \bibinfo {author} {\bibfnamefont {J.~F.}\ \bibnamefont
  {Fitzsimons}},\ and\ \bibinfo {author} {\bibfnamefont {S.~C.}\ \bibnamefont
  {Benjamin}},\ }\href@noop {} {\bibfield  {journal} {\bibinfo  {journal}
  {Physical Review X}\ }\textbf {\bibinfo {volume} {4}},\ \bibinfo {pages}
  {041041} (\bibinfo {year} {2014})}\BibitemShut {NoStop}%
\bibitem [{\citenamefont {Nigmatullin}\ \emph {et~al.}(2016)\citenamefont
  {Nigmatullin}, \citenamefont {Ballance}, \citenamefont {de~Beaudrap},\ and\
  \citenamefont {Benjamin}}]{nigmatullin2016minimally}%
  \BibitemOpen
  \bibfield  {author} {\bibinfo {author} {\bibfnamefont {R.}~\bibnamefont
  {Nigmatullin}}, \bibinfo {author} {\bibfnamefont {C.~J.}\ \bibnamefont
  {Ballance}}, \bibinfo {author} {\bibfnamefont {N.}~\bibnamefont
  {de~Beaudrap}},\ and\ \bibinfo {author} {\bibfnamefont {S.~C.}\ \bibnamefont
  {Benjamin}},\ }\href@noop {} {\bibfield  {journal} {\bibinfo  {journal} {New
  Journal of Physics}\ }\textbf {\bibinfo {volume} {18}},\ \bibinfo {pages}
  {103028} (\bibinfo {year} {2016})}\BibitemShut {NoStop}%
\bibitem [{\citenamefont {Krastanov}\ \emph {et~al.}(2019)\citenamefont
  {Krastanov}, \citenamefont {Albert},\ and\ \citenamefont
  {Jiang}}]{krastanov2019optimized}%
  \BibitemOpen
  \bibfield  {author} {\bibinfo {author} {\bibfnamefont {S.}~\bibnamefont
  {Krastanov}}, \bibinfo {author} {\bibfnamefont {V.~V.}\ \bibnamefont
  {Albert}},\ and\ \bibinfo {author} {\bibfnamefont {L.}~\bibnamefont
  {Jiang}},\ }\href@noop {} {\bibfield  {journal} {\bibinfo  {journal}
  {Quantum}\ }\textbf {\bibinfo {volume} {3}},\ \bibinfo {pages} {123}
  (\bibinfo {year} {2019})}\BibitemShut {NoStop}%
\bibitem [{\citenamefont {Jansen}\ \emph {et~al.}(2022)\citenamefont {Jansen},
  \citenamefont {Goodenough}, \citenamefont {de~Bone}, \citenamefont
  {Gijswijt},\ and\ \citenamefont {Elkouss}}]{jansen2022enumerating}%
  \BibitemOpen
  \bibfield  {author} {\bibinfo {author} {\bibfnamefont {S.}~\bibnamefont
  {Jansen}}, \bibinfo {author} {\bibfnamefont {K.}~\bibnamefont {Goodenough}},
  \bibinfo {author} {\bibfnamefont {S.}~\bibnamefont {de~Bone}}, \bibinfo
  {author} {\bibfnamefont {D.}~\bibnamefont {Gijswijt}},\ and\ \bibinfo
  {author} {\bibfnamefont {D.}~\bibnamefont {Elkouss}},\ }\href@noop {}
  {\bibfield  {journal} {\bibinfo  {journal} {Quantum}\ }\textbf {\bibinfo
  {volume} {6}},\ \bibinfo {pages} {715} (\bibinfo {year} {2022})}\BibitemShut
  {NoStop}%
\bibitem [{\citenamefont {Goodenough}\ \emph {et~al.}(2023)\citenamefont
  {Goodenough}, \citenamefont {de~Bone}, \citenamefont {Addala}, \citenamefont
  {Krastanov}, \citenamefont {Jansen}, \citenamefont {Gijswijt},\ and\
  \citenamefont {Elkouss}}]{goodenough2023near}%
  \BibitemOpen
  \bibfield  {author} {\bibinfo {author} {\bibfnamefont {K.}~\bibnamefont
  {Goodenough}}, \bibinfo {author} {\bibfnamefont {S.}~\bibnamefont {de~Bone}},
  \bibinfo {author} {\bibfnamefont {V.~L.}\ \bibnamefont {Addala}}, \bibinfo
  {author} {\bibfnamefont {S.}~\bibnamefont {Krastanov}}, \bibinfo {author}
  {\bibfnamefont {S.}~\bibnamefont {Jansen}}, \bibinfo {author} {\bibfnamefont
  {D.}~\bibnamefont {Gijswijt}},\ and\ \bibinfo {author} {\bibfnamefont
  {D.}~\bibnamefont {Elkouss}},\ }\href@noop {} {\bibfield  {journal} {\bibinfo
   {journal} {arXiv preprint arXiv:2303.11465}\ } (\bibinfo {year}
  {2023})}\BibitemShut {NoStop}%
\bibitem [{\citenamefont {Bennett}\ \emph
  {et~al.}(1996{\natexlab{b}})\citenamefont {Bennett}, \citenamefont
  {DiVincenzo}, \citenamefont {Smolin},\ and\ \citenamefont
  {Wootters}}]{bennett1996mixed}%
  \BibitemOpen
  \bibfield  {author} {\bibinfo {author} {\bibfnamefont {C.~H.}\ \bibnamefont
  {Bennett}}, \bibinfo {author} {\bibfnamefont {D.~P.}\ \bibnamefont
  {DiVincenzo}}, \bibinfo {author} {\bibfnamefont {J.~A.}\ \bibnamefont
  {Smolin}},\ and\ \bibinfo {author} {\bibfnamefont {W.~K.}\ \bibnamefont
  {Wootters}},\ }\href@noop {} {\bibfield  {journal} {\bibinfo  {journal}
  {Phys. Rev. A}\ }\textbf {\bibinfo {volume} {54}},\ \bibinfo {pages} {3824}
  (\bibinfo {year} {1996}{\natexlab{b}})}\BibitemShut {NoStop}%
\bibitem [{\citenamefont {Muralidharan}\ \emph {et~al.}(2016)\citenamefont
  {Muralidharan}, \citenamefont {Li}, \citenamefont {Kim}, \citenamefont
  {L{\"u}tkenhaus}, \citenamefont {Lukin},\ and\ \citenamefont
  {Jiang}}]{muralidharan2016optimal}%
  \BibitemOpen
  \bibfield  {author} {\bibinfo {author} {\bibfnamefont {S.}~\bibnamefont
  {Muralidharan}}, \bibinfo {author} {\bibfnamefont {L.}~\bibnamefont {Li}},
  \bibinfo {author} {\bibfnamefont {J.}~\bibnamefont {Kim}}, \bibinfo {author}
  {\bibfnamefont {N.}~\bibnamefont {L{\"u}tkenhaus}}, \bibinfo {author}
  {\bibfnamefont {M.~D.}\ \bibnamefont {Lukin}},\ and\ \bibinfo {author}
  {\bibfnamefont {L.}~\bibnamefont {Jiang}},\ }\href@noop {} {\bibfield
  {journal} {\bibinfo  {journal} {Scientific reports}\ }\textbf {\bibinfo
  {volume} {6}},\ \bibinfo {pages} {20463} (\bibinfo {year}
  {2016})}\BibitemShut {NoStop}%
\bibitem [{\citenamefont {Laflamme}\ \emph {et~al.}(1996)\citenamefont
  {Laflamme}, \citenamefont {Miquel}, \citenamefont {Paz},\ and\ \citenamefont
  {Zurek}}]{laflamme1996perfect}%
  \BibitemOpen
  \bibfield  {author} {\bibinfo {author} {\bibfnamefont {R.}~\bibnamefont
  {Laflamme}}, \bibinfo {author} {\bibfnamefont {C.}~\bibnamefont {Miquel}},
  \bibinfo {author} {\bibfnamefont {J.~P.}\ \bibnamefont {Paz}},\ and\ \bibinfo
  {author} {\bibfnamefont {W.~H.}\ \bibnamefont {Zurek}},\ }\href@noop {}
  {\bibfield  {journal} {\bibinfo  {journal} {Phys. Rev. Lett.}\ }\textbf
  {\bibinfo {volume} {77}},\ \bibinfo {pages} {198} (\bibinfo {year}
  {1996})}\BibitemShut {NoStop}%
\bibitem [{\citenamefont {Gottesman}(1997)}]{gottesman_thesis}%
  \BibitemOpen
  \bibfield  {author} {\bibinfo {author} {\bibfnamefont {D.}~\bibnamefont
  {Gottesman}},\ }\href@noop {} {\bibfield  {journal} {\bibinfo  {journal}
  {arXiv preprint quant-ph/9705052}\ } (\bibinfo {year} {1997})}\BibitemShut
  {NoStop}%
\bibitem [{\citenamefont {Gottesman}(1998)}]{gottesman1998heisenberg}%
  \BibitemOpen
  \bibfield  {author} {\bibinfo {author} {\bibfnamefont {D.}~\bibnamefont
  {Gottesman}},\ }\href@noop {} {\bibfield  {journal} {\bibinfo  {journal}
  {arXiv preprint quant-ph/9807006}\ } (\bibinfo {year} {1998})}\BibitemShut
  {NoStop}%
\bibitem [{\citenamefont {Aaronson}\ and\ \citenamefont
  {Gottesman}(2004)}]{aaronson2004improved}%
  \BibitemOpen
  \bibfield  {author} {\bibinfo {author} {\bibfnamefont {S.}~\bibnamefont
  {Aaronson}}\ and\ \bibinfo {author} {\bibfnamefont {D.}~\bibnamefont
  {Gottesman}},\ }\href@noop {} {\bibfield  {journal} {\bibinfo  {journal}
  {Physical Review A}\ }\textbf {\bibinfo {volume} {70}},\ \bibinfo {pages}
  {052328} (\bibinfo {year} {2004})}\BibitemShut {NoStop}%
\bibitem [{\citenamefont {Anders}\ and\ \citenamefont
  {Briegel}(2006)}]{anders06}%
  \BibitemOpen
  \bibfield  {author} {\bibinfo {author} {\bibfnamefont {S.}~\bibnamefont
  {Anders}}\ and\ \bibinfo {author} {\bibfnamefont {H.~J.}\ \bibnamefont
  {Briegel}},\ }\href@noop {} {\bibfield  {journal} {\bibinfo  {journal}
  {Physical Review A}\ }\textbf {\bibinfo {volume} {73}} (\bibinfo {year}
  {2006})}\BibitemShut {NoStop}%
\bibitem [{\citenamefont {Dehaene}\ \emph {et~al.}(2003)\citenamefont
  {Dehaene}, \citenamefont {Van~den Nest}, \citenamefont {De~Moor},\ and\
  \citenamefont {Verstraete}}]{dehaene2003local}%
  \BibitemOpen
  \bibfield  {author} {\bibinfo {author} {\bibfnamefont {J.}~\bibnamefont
  {Dehaene}}, \bibinfo {author} {\bibfnamefont {M.}~\bibnamefont {Van~den
  Nest}}, \bibinfo {author} {\bibfnamefont {B.}~\bibnamefont {De~Moor}},\ and\
  \bibinfo {author} {\bibfnamefont {F.}~\bibnamefont {Verstraete}},\
  }\href@noop {} {\bibfield  {journal} {\bibinfo  {journal} {Physical Review
  A}\ }\textbf {\bibinfo {volume} {67}},\ \bibinfo {pages} {022310} (\bibinfo
  {year} {2003})}\BibitemShut {NoStop}%
\bibitem [{\citenamefont {Koenig}\ and\ \citenamefont
  {Smolin}(2014)}]{koenig2014efficiently}%
  \BibitemOpen
  \bibfield  {author} {\bibinfo {author} {\bibfnamefont {R.}~\bibnamefont
  {Koenig}}\ and\ \bibinfo {author} {\bibfnamefont {J.~A.}\ \bibnamefont
  {Smolin}},\ }\href@noop {} {\bibfield  {journal} {\bibinfo  {journal}
  {Journal of Mathematical Physics}\ }\textbf {\bibinfo {volume} {55}},\
  \bibinfo {pages} {122202} (\bibinfo {year} {2014})}\BibitemShut {NoStop}%
\bibitem [{\citenamefont {Krastanov}\ and\ \citenamefont
  {ShuGe-MIT}(2023)}]{stefan_krastanov_2023_8137209}%
  \BibitemOpen
  \bibfield  {author} {\bibinfo {author} {\bibfnamefont {S.}~\bibnamefont
  {Krastanov}}\ and\ \bibinfo {author} {\bibnamefont {ShuGe-MIT}},\ }\href
  {https://doi.org/10.5281/zenodo.8137209} {\bibinfo {title}
  {Quantumsavory/bpgates.jl}} (\bibinfo {year} {2023})\BibitemShut {NoStop}%
\bibitem [{\citenamefont {Addala}\ and\ \citenamefont
  {Krastanov}(2023)}]{vaishnavi_addala_2023_7844179}%
  \BibitemOpen
  \bibfield  {author} {\bibinfo {author} {\bibfnamefont {V.}~\bibnamefont
  {Addala}}\ and\ \bibinfo {author} {\bibfnamefont {S.}~\bibnamefont
  {Krastanov}},\ }\href {https://doi.org/10.5281/zenodo.7844179} {\bibinfo
  {title} {qevo\_optimizer}} (\bibinfo {year} {2023})\BibitemShut {NoStop}%
\bibitem [{\citenamefont {Ozols}(2008)}]{ozols08}%
  \BibitemOpen
  \bibfield  {author} {\bibinfo {author} {\bibfnamefont {M.}~\bibnamefont
  {Ozols}},\ }\href@noop {} {\bibfield  {journal} {\bibinfo  {journal} {Essays
  at University of Waterloo, Spring}\ } (\bibinfo {year} {2008})}\BibitemShut
  {NoStop}%
\bibitem [{\citenamefont {Garcia}\ \emph {et~al.}(2012)\citenamefont {Garcia},
  \citenamefont {Markov},\ and\ \citenamefont {Cross}}]{garcia12}%
  \BibitemOpen
  \bibfield  {author} {\bibinfo {author} {\bibfnamefont {H.~J.}\ \bibnamefont
  {Garcia}}, \bibinfo {author} {\bibfnamefont {I.~L.}\ \bibnamefont {Markov}},\
  and\ \bibinfo {author} {\bibfnamefont {A.~W.}\ \bibnamefont {Cross}},\
  }\href@noop {} {\bibfield  {journal} {\bibinfo  {journal} {arXiv preprint
  arXiv:1210.6646}\ } (\bibinfo {year} {2012})}\BibitemShut {NoStop}%
\end{thebibliography}%

\newpage
\pagebreak
\newpage
\pagebreak

\appendix

\section*{Appendix}

In this appendix, we provide a few more parameter sweeps for the purification circuits we have generated. Then we give a self-contained introduction to the basics of entanglement purification and Clifford circuit simulation.

\subsection{Mutual Information in Cost Function}

We introduce a formalization of correlation for qubits that we used in our study which is mutual information, $I$. $I(X;Y) = H(X) + H(Y) - H(X;Y)$ where $X$ and $Y$ are quantum systems, $H(X)$ and $H(Y)$ are their marginal entropies, and $H(X;Y)$ is the joint entropy. $I$ can be thought of intuitively as how much one system tells us about the other.

\begin{figure}
    \centering
    \includegraphics[width=\linewidth]{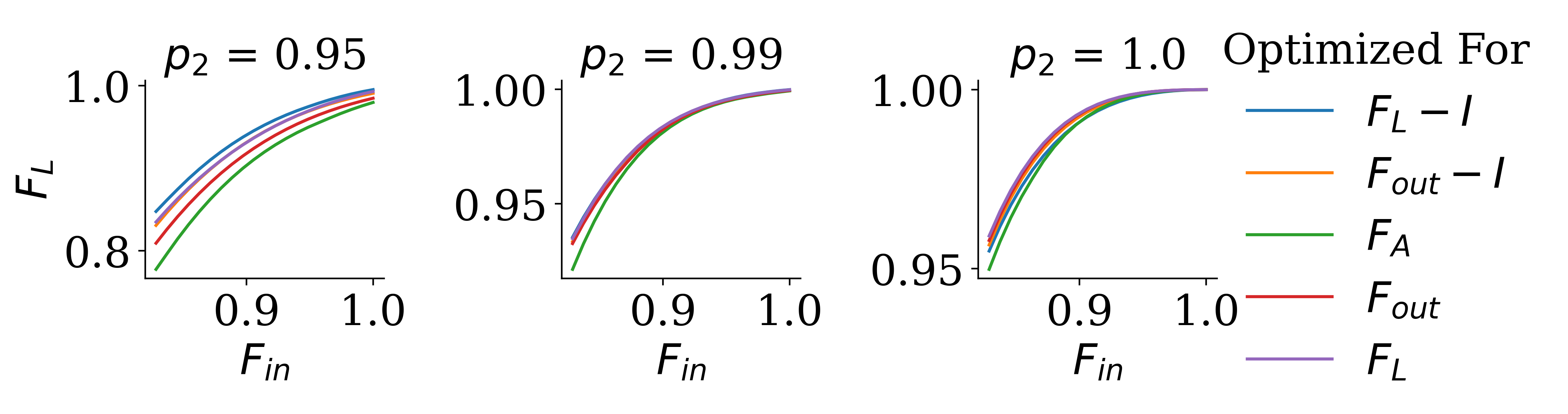}
    \caption{\textbf{Best teleportation of logical qubits}. The plot is $F_\textrm{in}$ vs $F_L$ (the fidelity of the teleported qubit followed by error correction). The best $F_L$ for each of the hardware parameters is shown. Unlike the corresponding plot from the main text, here every single point shows circuits optimized for the corresponding $F_\textrm{in}$ and $p_2$, instead of having one single set of circuits optimized for a fixed pair of hardware parameters and then evaluated at hardware parameters they were not optimized for. This plot is provided for completeness and it does not change the conclusions of the main text.}
    \label{fig:TeleportedBest}
\end{figure}

\begin{figure}
    \centering
    \includegraphics[width=\linewidth]{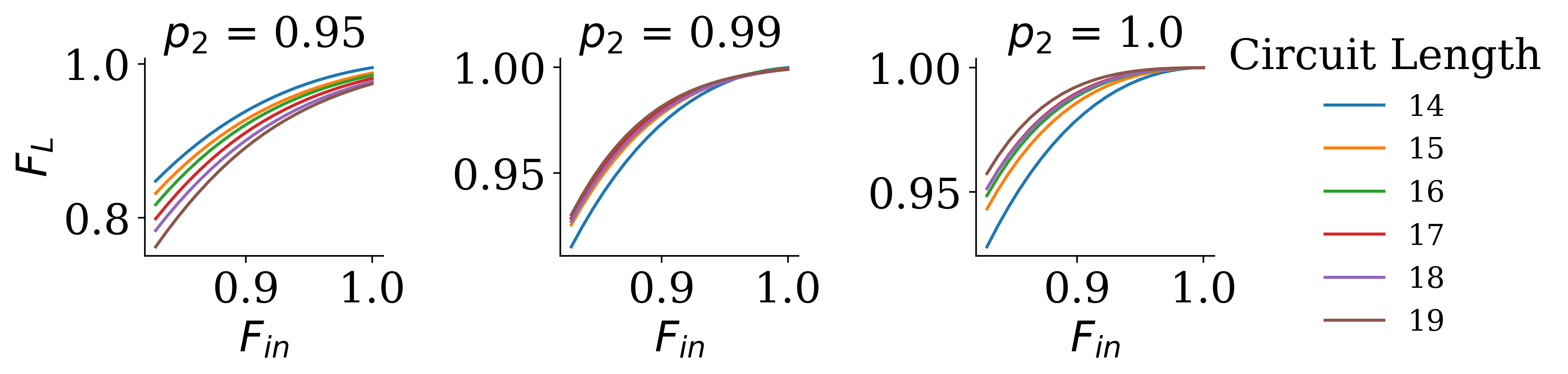}
    \caption{\textbf{Circuit length as a factor in fidelity}. The plot is $F_\textrm{in}$ vs $F_L$ and we show the circuits from Fig.~\ref{fig:Teleportation} that were optimized for $F_L$ and $F_L - I$. We provide this plot in order to better visualize the effects of circuit length (number of bilateral gates), which was not visible in the plot in the main text.}
    \label{fig:CircuitLength}
\end{figure}

We will compare the circuits generated with cost functions $F_L$ and $F_L - I$ seen in Fig.~\ref{fig:Teleportation}. The circuits optimized for $F_L$ were on average one gate longer. The two sets of circuits perform very similarly at the gate fidelity they were optimized for, $p_2=0.99$. Therefore, a larger proportion of the error in the longer circuits is because of gate error. This leads to better performance for longer circuits at higher gate fidelities and worse performance at lower gate fidelities. A visualization of $F_L$ with respect to circuit length is shown in Fig.~\ref{fig:CircuitLength} and Fig.~\ref{fig:TeleportedBest}. For these circuits, perturbing them (while keeping $F_L$ comparable) by adding a CNOT leads to increased $I$ while deleting a CNOT leads to decreased $I$. We know CNOTs add in correlations between errors on different qubits so we conjecture that penalizing $I$ leads to finding circuits with a shorter length in order to avoid some correlations. We note that circuits tend to have higher $I$ when evaluated at higher $p_2$ and/or lower $F_\textrm{in}$. We think this is because the errors that become correlated due to CNOTs are much more likely at lower $F_\textrm{in}$ and the errors that happen due to the gates are at least an order of magnitude less likely and serve as noise between the correlations of the errors due to $F_\textrm{in}$. This is supported by what we found when optimizing circuits for lower $p_2$. In this case, the circuits optimized for $F_L - I$ were on average much closer in length to those optimized for $F_L$. These circuits additionally had much smaller $I$, when evaluated over a range of $p_2$, than those optimized for higher $p_2$. 

Overall, we find that $I$ can be used as a regularizing function for finding circuits that will perform better at lower gate fidelities.

\subsection{Clifford Group}

On $N$ qubits, the set of Pauli matrices is defined as:
\begin{equation}\label{eq:two}
    P_N=\{\sigma_1\otimes\dots\otimes\sigma_N|\sigma_i\in \{I,X,Y,Z\}\}
\end{equation}

The group $P_N/U(1)$ is isomorphic to a vector space over $\mathbb{F}_2$ with dimension $2N$ via identification where the multiplication of Pauli matrices corresponds to the addition of vectors \cite{ozols08}.

To define the Clifford group, consider $P_N^*=P_N\backslash \{I^{\otimes N}\}$. Each matrix has eigenvalues of $\pm 1$ with equal multiplicity. Then the group of $N$-qubit Clifford gates is defined as
\begin{eqnarray}
    \mathcal{C}_N = \{U\in U(2^N)|U\pm P_NU^*=\pm P_N\} / U(1)
\end{eqnarray}
where $U(N)$ is the unitary group and $U(1)$ is the circle group, consisting of all complex numbers with absolute value 1. Here, by taking the quotient over $U(1)$, we ignore the global phases. The Clifford group can also be interpreted as the normalizer of the $n$-qubit Pauli group \cite{ozols08}.

For any Clifford gate $U\in \mathcal{C}_N$, it is enough to specify what's the state that $X_i$ and $Z_i$, where $i=1,2,\dots,N$, are mapped to. This is because $X_i$ and $Z_i$, where $i=1,2,\dots,N$, form a basis of the vector space. Images of all $X$'s and $Z$'s, except for $X_i$ and $Z_i$, should commute under conjugation by a Clifford gate. For example, for $\mathcal{C}_1$, $X$ can be mapped to any $\pm P_1^*$, but the image of $Z$ must anti-commute with the image of $X$. Hence, the total size of $\mathcal{C}_1$ is $6\times 4 = 24$ \cite{ozols08}.

Similarly, given $\mathcal{C}_{N-1}$, we can compute $\mathcal{C}_N$ by considering where $X_N, Z_N$ are mapped to. There are $|\pm P_N^*|=2(4^N-1)$ ways to map $X_N$, which will anti-commute with half of $P_N$. Hence, there are $2(4^N-1)4^N$ ways to map $X_N$ and $Z_N$ after we get $\mathcal{C}_{N-1}$ \cite{ozols08}. Therefore,
\begin{equation}
    |\mathcal{C}_N|=2(4^{N}-1)4^{N}|\mathcal{C}_{N-1}|.
\end{equation}

We can also solve this recurrence relationship to get 
\begin{equation}\label{eq:size}
    |\mathcal{C}_N|=\prod_{j=1}^N 2(4^j-1)4^j=2^{N^2+2N}\prod_{j=1}^N (4^j-1)
\end{equation}

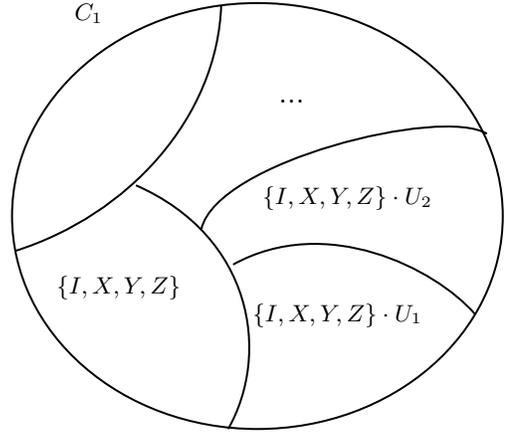
\begin{figure}
\tikzset{every picture/.style={line width=0.75pt}} 

\begin{tikzpicture}[x=0.75pt,y=0.75pt,yscale=-1,xscale=1]

\draw   (256,145.5) .. controls (256,86.13) and (311.4,38) .. (379.75,38) .. controls (448.1,38) and (503.5,86.13) .. (503.5,145.5) .. controls (503.5,204.87) and (448.1,253) .. (379.75,253) .. controls (311.4,253) and (256,204.87) .. (256,145.5) -- cycle ;
\draw    (258,163) .. controls (314.5,145) and (359.5,96) .. (361.5,39) ;
\draw    (365,253) .. controls (386.5,215) and (377.5,156) .. (318.5,130) ;
\draw    (489.5,195) .. controls (469.5,172) and (414.5,144) .. (367.5,170) ;
\draw    (495.5,104) .. controls (472.5,90) and (359.5,118) .. (351.5,152) ;

\draw (389,85) node [anchor=north west][inner sep=0.75pt]  [font=\large] [align=left] {...};
\draw (277,174) node [anchor=north west][inner sep=0.75pt]   [align=left] {$\displaystyle \{I,X,Y,Z\}$};
\draw (376,188) node [anchor=north west][inner sep=0.75pt]   [align=left] {$\displaystyle \{I,X,Y,Z\} \cdot U_{1}$};
\draw (286,37) node [anchor=north west][inner sep=0.75pt]   [align=left] {$\displaystyle C_{1}$};
\draw (381,129) node [anchor=north west][inner sep=0.75pt]   [align=left] {$\displaystyle \{I,X,Y,Z\} \cdot U_{2}$};

\end{tikzpicture}
\caption{\label{fig:c1} \textbf{Graphical illustration of $C_1^*=C_1/\mathbb{Z}_2^2$}. }
\end{figure}

Notice that in Eq. 1, we have shown that Bell states can be converted to each other using Pauli matrices. Hence, applying $\{I, X, Y, Z\}$ or not does not matter in our consideration of Bell preserving gates since these gates only permute between Bell basis. For example, for $\mathcal{C}_1$ as shown in Fig.~\ref{fig:c1}, all $\{I, X, Y, Z\}$ are in the same coset. All $P_1=\{I\cdot U, X\cdot U, Y\cdot U, Z\cdot U\}$ are in the same coset since the phases of the Bell states are absorbed into the cosets $P_1$ such that the elements in the quotient group are phaseless. We want to work with this quotient group since we are only concerned about circuits that preserve the Bell-diagonal state irrespective of the possible change in the phases. Hence, we want to take the quotient group of $\mathcal{C}_N$ with respect to $\mathbb{Z}_2^{2N}
$, where we ignore the phases of the elements of the Pauli group, i.e.
\begin{align}
    \mathcal{C}_N^* &= \{U\in U(2^N)|U\mathcal{P}_NU^*
    =\mathcal{P}_N\}/U(1)&\\
    \mathcal{C}_N^* &= \mathcal{C}_N / \mathbb{Z}_2^{2N}.
\end{align}

Similarly, we can find the recurrence relation for the size of $\mathcal{C}^*_N$

\begin{equation}
    |\mathcal{C}^*_N|=(4^{N}-1)4^{N}/2|\mathcal{C}^*_{N-1}|.
\end{equation}
    We can solve this recurrence to get
\begin{equation}
    |\mathcal{C}^*_N|=\prod_{j=1}^N (4^j-1)4^j/2=2^{N^2}\prod_{j=1}^N (4^j-1).
\end{equation}
    Compared to Eq.~\ref{eq:size}, we can see that the reason why we want to consider gates ignoring global phases is that $\mathcal{C}^*_N$ is exponentially smaller than $\mathcal{C}_N$. If we want to restore the full group of gates, we can just add phases afterward.

\subsection{\label{sec:stabgroup}Stabilizer Group}
A unitary operator $U$ stabilizers a state $|\psi\rangle$ if $U|\psi\rangle=|\psi\rangle$. For a given $N$-qubit pure state $|\psi\rangle$, the stabilizer group $S(|\psi\rangle)=\langle S_1, ..., S_k\rangle$ with $k\leq N$ is an Abelian group of $\pm P_N$. We can also write the $N\times N$ stabilizer matrix for a state, whose rows are all stabilizers. 
\begin{equation}\label{eq:stabsize}
    S(|\psi\rangle)\cong \mathbb{Z}^k_2.
\end{equation}

If $k=N$, $|\psi\rangle$ is uniquely specified by the stabilizer group and is called a stabilizer state \cite{garcia12}. Not all quantum states can be represented in a stabilizer form. Only Pauli-stabilized states can be represented in stabilizer form. Any such state will be the common eigenvector of all stabilizers with an eigenvalue equal to 1 \cite{garcia12}.

Stabilizers can be efficiently represented on classical computers using binary matrices or tableau, with space $\mathcal{O}(N^2)$ for an $n$-qubit pure state. Since $P_1\cong \mathbb{Z}_2^2$, the Pauli matrices can be represented using only two bits, e.g. $I=00,\ Z=01,\ X=10,\ Y=11$. Therefore, the stabilizer matrix can be represented as an $N\times 2N$ tableau \cite{garcia12}. 

\subsection{\label{sec:canonicalization}Canonicalization of Stabilizer Matrices}
Although the state is uniquely determined by the stabilizer group, the set of generators is not unique. Hence, in order to determine whether two stabilizer groups or stabilizer matrices represent the same state, we need a canonical form of the stabilizer matrices for direct comparison \cite{garcia12}.

Now consider the set of generators of the stabilizer group $S(|\psi\rangle)$. By Eq.~\ref{eq:stabsize}, each generator imposes a linear constraint on $|\psi\rangle$ that divides the Hilbert space into half, so the set of generators can be viewed as a system of linear equations whose solution yields the $2^N$
 basis amplitudes that make up $|\psi\rangle$. Hence, we can use Gaussian elimination to obtain the basis amplitudes from a generator set \cite{garcia12}.
 
Any stabilizer can be rearranged by applying a sequence of elementary row operations, including transposition and multiplication, which do not modify the stabilizer state. Similar to Gauss-Jordan elimination, the algorithm proposed in \cite{garcia12} rearranges the stabilizer state into a \textit{row-reduced echelon form} that contains a minimum set of generators with $X$ and $Y$ literals appearing at the top and generators containing a minimum set of $Z$ literals appearing at the bottom of the matrix.

\subsection{\label{sec:entanglement}Entanglement Purification Circuit}

In an entanglement purification protocol, Alice and Bob start with a small number of entanglements with low fidelity. They perform a series of local Clifford operations and measurements and use classical communications to determine if the protocol is successful. They will be able to obtain a single pair of higher fidelity.

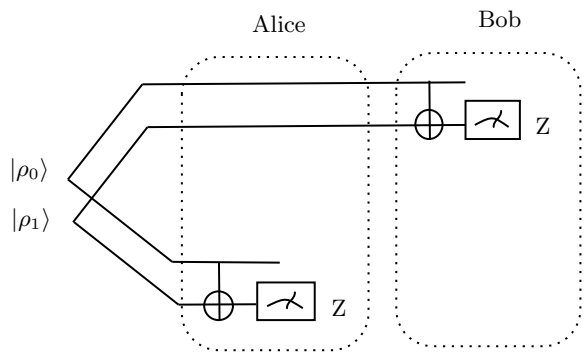
\begin{figure}
\tikzset{every picture/.style={line width=0.75pt}} 

\begin{tikzpicture}[x=0.75pt,y=0.75pt,yscale=-1,xscale=1]

\draw    (228.53,97) -- (281.15,138.6) ;
\draw    (266.07,49) -- (228.53,97) ;
\draw    (281.15,138.6) -- (335.39,139) ;
\draw    (231.67,118.6) -- (284.29,160.2) ;
\draw    (268.66,70.6) -- (231.12,118.6) ;
\draw    (284.29,160.2) -- (323.5,160) ;
\draw    (304.71,139) -- (305,168) ;
\draw   (297.25,160.75) .. controls (297.23,156.75) and (300.45,153.5) .. (304.46,153.5) .. controls (308.46,153.5) and (311.73,156.75) .. (311.75,160.75) .. controls (311.77,164.75) and (308.55,168) .. (304.54,168) .. controls (300.54,168) and (297.27,164.75) .. (297.25,160.75) -- cycle ;
\draw   (324,149) -- (353,149) -- (353,168) -- (324,168) -- cycle ;
\draw    (329,161) .. controls (339,156) and (339,157) .. (347,162) ;
\draw  [line width=0.75] [line join = round][line cap = round] (339,162) .. controls (339,158.35) and (343,157.79) .. (343,154) ;
\draw    (266.07,49) -- (429.01,48) ;
\draw    (268.66,70.6) -- (429.06,69.5) ;
\draw    (410.82,47.17) -- (411.1,76.6) ;
\draw   (403.77,69.24) .. controls (403.75,65.18) and (406.8,61.89) .. (410.59,61.89) .. controls (414.37,61.89) and (417.46,65.18) .. (417.48,69.24) .. controls (417.5,73.3) and (414.45,76.6) .. (410.66,76.6) .. controls (406.88,76.6) and (403.79,73.3) .. (403.77,69.24) -- cycle ;
\draw   (429.06,57.32) -- (456.47,57.32) -- (456.47,76.6) -- (429.06,76.6) -- cycle ;
\draw    (433.78,69.5) .. controls (443.23,64.42) and (443.23,65.44) .. (450.8,70.51) ;
\draw  [line width=0.75] [line join = round][line cap = round] (443.23,70.51) .. controls (443.23,66.81) and (447.02,66.24) .. (447.02,62.39) ;
\draw  [dash pattern={on 0.84pt off 2.51pt}] (286,54.03) .. controls (286,43.63) and (294.43,35.2) .. (304.83,35.2) -- (361.31,35.2) .. controls (371.71,35.2) and (380.14,43.63) .. (380.14,54.03) -- (380.14,164.37) .. controls (380.14,174.77) and (371.71,183.2) .. (361.31,183.2) -- (304.83,183.2) .. controls (294.43,183.2) and (286,174.77) .. (286,164.37) -- cycle ;
\draw  [dash pattern={on 0.84pt off 2.51pt}] (394.14,51.77) .. controls (394.14,41.52) and (402.45,33.2) .. (412.71,33.2) -- (468.43,33.2) .. controls (478.68,33.2) and (487,41.52) .. (487,51.77) -- (487,162.63) .. controls (487,172.88) and (478.68,181.2) .. (468.43,181.2) -- (412.71,181.2) .. controls (402.45,181.2) and (394.14,172.88) .. (394.14,162.63) -- cycle ;

\draw (360,156) node [anchor=north west][inner sep=0.75pt]   [align=left] {Z};
\draw (462.78,64.55) node [anchor=north west][inner sep=0.75pt]   [align=left] {Z};
\draw (320,13) node [anchor=north west][inner sep=0.75pt]   [align=left] {Alice};
\draw (434,10) node [anchor=north west][inner sep=0.75pt]   [align=left] {Bob};
\draw (197.57,85) node [anchor=north west][inner sep=0.75pt]   [align=left] {$\displaystyle |\rho _{0} \rangle $};
\draw (198.46,110) node [anchor=north west][inner sep=0.75pt]   [align=left] {$\displaystyle |\rho _{1} \rangle $};

\end{tikzpicture}
\caption{\label{fig:examplecircuit} \textbf{An example of a purification circuit}. This circuit purifies a Bell pair between Alice and Bob using a sacrificial pair.}
\end{figure}

For example, in Fig. ~\ref{fig:examplecircuit}, Alice and Bob share two Bell pairs and sacrifice one of them to get one pair with higher fidelity. When Alice and Bob perform local CNOT gate, any possible error in the upper qubit will be propagated to the lower qubit. When they perform a coincidence Z measurement on the lower pair, if it has correlated results, we accept the upper pair. Otherwise, an error is detected and we reject the upper pair. If we interpret a purification circuit as an error-detecting circuit, we have three types of errors: X is a bit flip error, Z is a phase flip error, and Y is a combination of both. Coincidence measurement can only rule out half of the states but it can't exactly pin down the error. For example, in the circuit in Fig.~\ref{fig:examplecircuit}, Alice and Bob take a Z coincidence measurement. The Bell states $|A\rangle$ and $|D\rangle$ will give correlated measurement results. An error is flagged if either $|B\rangle$ or $|C\rangle$ is measured, so only X and Y errors can be detected by this circuit. The fidelity is thus $p_I/(p_I+p_Z)$. Since $(p_I+p_Z)\leq1$, the fidelity is improved.

\end{document}